\documentclass[a4paper,fleqn,usenatbib]{mnras}

\usepackage{newtxtext,newtxmath}

\usepackage[T1]{fontenc}
\usepackage{ae,aecompl}

\usepackage{graphicx}	
\usepackage{amsmath}	
\usepackage{amssymb}	
\usepackage{subfig}

\usepackage{calrsfs}

\title[ADP--IS method for impact probability computation]{A differential algebra based importance sampling method for impact probability computation on Earth resonant returns of Near Earth Objects}

\author[M. Losacco et al.]{
Matteo Losacco,$^{1}$\thanks{E-mail: matteo.losacco@polimi.it}
Pierluigi Di Lizia,$^{1}$
Roberto Armellin$^{2}$
and Alexander Wittig$^{3}$
\\
$^{1}$Department of Aerospace Science and Technology, Politecnico di Milano, Via G. La Masa 34, 20156 Milano, Italy\\
$^{2}$Surrey Space Centre, University of Surrey, GU2 7XH Guildford, United Kingdom\\
$^{3}$Aeronautics, Astronautics and Computational Engineering Unit, University of Southampton, SO17 1BJ, Southampton, United Kingdom
}

\date{Accepted XXX. Received YYY; in original form ZZZ}

\pubyear{2017}

\begin{document}
\label{firstpage}
\pagerange{\pageref{firstpage}--\pageref{lastpage}}
\maketitle

\begin{abstract}
A differential algebra based importance sampling method for uncertainty propagation and impact probability computation on the first resonant returns of Near Earth Objects is presented in this paper. Starting from the results of an orbit determination process, we use a differential algebra based automatic domain pruning to estimate resonances and automatically propagate in time the regions of the initial uncertainty set that include the resonant return of interest. The result is a list of polynomial state vectors, each mapping specific regions of the uncertainty set from the observation epoch to the resonant return. Then, we employ a Monte Carlo importance sampling technique on the generated subsets for impact probability computation. We assess the performance of the proposed approach on the case of asteroid (99942) Apophis. A sensitivity analysis on the main parameters of the technique is carried out, providing guidelines for their selection. We finally compare the results of the proposed method to standard and advanced orbital sampling techniques.
\end{abstract}




\section{Introduction}

Over the last thirty years, significant efforts have been devoted to develop new tools for detection and prediction of planetary encounters and potential impacts by Near Earth Objects (NEO). The task introduces relevant challenges due to the imperative of early detection and accurate estimation and propagation of their state and associated uncertainty set \citep{Chesley2005}. The problem is made more complicated by the fact that the dynamics describing the motion of these objects is highly nonlinear, especially during close encounters with major bodies.
Nonlinearities of the orbital dynamics tend to significantly stretch the initial uncertainty sets during the time propagation. Nonlinearities are not confined to object dynamics only: even simple conversions between coordinate systems introduce nonlinearities, thus affecting the accuracy of classical propagation techniques \citep{Wittig2015}. Present day approaches for robust detection and prediction of planetary encounters and potential impacts by NEO mainly refer to linearised models or full nonlinear orbital sampling \citep{Farnocchia2015}. The impact probability computation by means of linear methods in the impact plane was introduced by \citet{Chodas1993}, whereas the introduction of the Monte Carlo technique to this problem was developed by \citet{Yeomans1994} and \citet{Chodas1999a}, who suggested to apply the method to sample the linear six dimensional confidence region at the observation epoch and then numerically integrate over the time interval of investigation using fully nonlinear equations \citep{Milani2002}. \citet{Milani1999}, \citet{Milani1999a} and \citet{Milani2000,Milani2000a} applied the multiple solutions approach to sample the central Line of Variations (LOV) of the nonlinear confidence region at the initial epoch and then numerically integrate over the time span of interest in a similar way. Within the framework of the impact probability computation of resonant returns, a well-known approach relies on the concept of keyholes, small regions of the impact plane of a specific close encounter such that, if an asteroid passes through one of them, it will hit the Earth on subsequent return \citep{Gronchi2001,Milani2002,Valsecchi2003}.

The preferred approach to detecting potential impacts depends on the uncertainty in the estimated orbit, the investigated time window and the dynamics between the observation epoch and the epoch of the expected impact \citep{Farnocchia2015}. Linear methods are preferred when linear approximations are reliable for both the orbit determination and uncertainty propagation. When these assumptions are not valid, one must resort to more computationally intensive techniques: among these, Monte Carlo methods are the most accurate but also the most computationally intensive, whereas the LOV method guarantees compute times 3-4 orders of magnitude lower than those required in MC simulations, though the LOV analysis may grow quite complex after it has been stretched and folded by multiple close planetary encounters, leaving open the possibility of missing some pathological cases \citep{Farnocchia2015}.

Alternative approaches rely on the use of Differential Algebra (DA). Differential algebra supplies the tools to compute the derivatives of functions within a computer environment, i.e. it provides the Taylor expansion of the flow of Ordinary Differential Equations (ODEs) by carrying out all the operations of any explicit integration scheme in the DA framework \citep{Berz1999,Wittig2015}. DA has already proven its efficiency in the nonlinear propagation of uncertainties \citep{Armellin2010, Morselli2012,Valli2013}. Nonetheless, the accuracy of the method drastically decreases in highly nonlinear dynamics. The propagation of asteroids motion after a close encounter with a major body is a typical case.

A DA based automatic domain splitting algorithm was presented by the authors in the past to overcome the limitations of simple DA propagation \citep{Wittig2014,Wittig2014a,Wittig2015}. The method can accurately propagate large sets of uncertainties in highly nonlinear dynamics and long term time spans. The propagation algorithm automatically splits the initial uncertainty domain into subsets when the polynomial expansions representing the current state do not meet predefined accuracy requirements. The performance of the algorithm was assessed on the case of asteroid (99942) Apophis, providing a description of the evolution of the uncertainty set to the epoch of predicted close encounters with Earth in 2036 and 2037 \citep{Wittig2015}. Though representing a significant improvement with respect to simple DA propagation, the approach required a not negligible computational effort in propagating the whole set of generated subdomains. Moreover, no information about the impact probability for asteroid Apophis was provided, as the propagation of the uncertainty set was stopped before the close encounters.

We present in this paper an evolution of the automatic domain splitting algorithm. The method, referred to as automatic domain pruning, automatically identifies possible resonances after a close encounter with a major body. Then, assuming no intervening close approaches with other celestial bodies in between, it optimizes the propagation to the first resonant returns, by limiting the propagation of the uncertainty set to the regions that generate a close encounter with that celestial body at the investigated epoch. The result is a list of polynomial state vectors, each mapping only specific subsets of the initial domain to the resonant return epoch. Taking advantage of the availability of the polynomial maps, a DA based Monte Carlo importance sampling  technique is then used to generate samples in the propagated subsets and provide an estimate for the impact probability at the epoch of the selected resonant return. The proposed approach does not apply any simplification step on the uncertainty domain associated with the orbit determination process. Thus, the method is proposed as an alternative approach with respect to equivalent techniques, such as a full Monte Carlo simulation or other six dimensional-based orbital sampling techniques, which will represent the main term of comparison for our analysis.

The paper is organized as follows. First, we present a description of the automatic domain pruning and importance sampling techniques, showing the application to the case of the first resonant return. Then, we apply the method to the critical case of asteroid (99942) Apophis, providing an estimate of the impact probability for the resonant return in 2036. Finally, we carry out a sensitivity analysis on the main parameters of the method, presenting a comparison with standard and advanced orbital sampling techniques.

\section{Differential Algebra and Automatic Domain Splitting}
\label{Section_ADS}

\begin{figure}
	\includegraphics[trim=0cm 0cm 0cm 0cm, clip=true, width=\columnwidth]{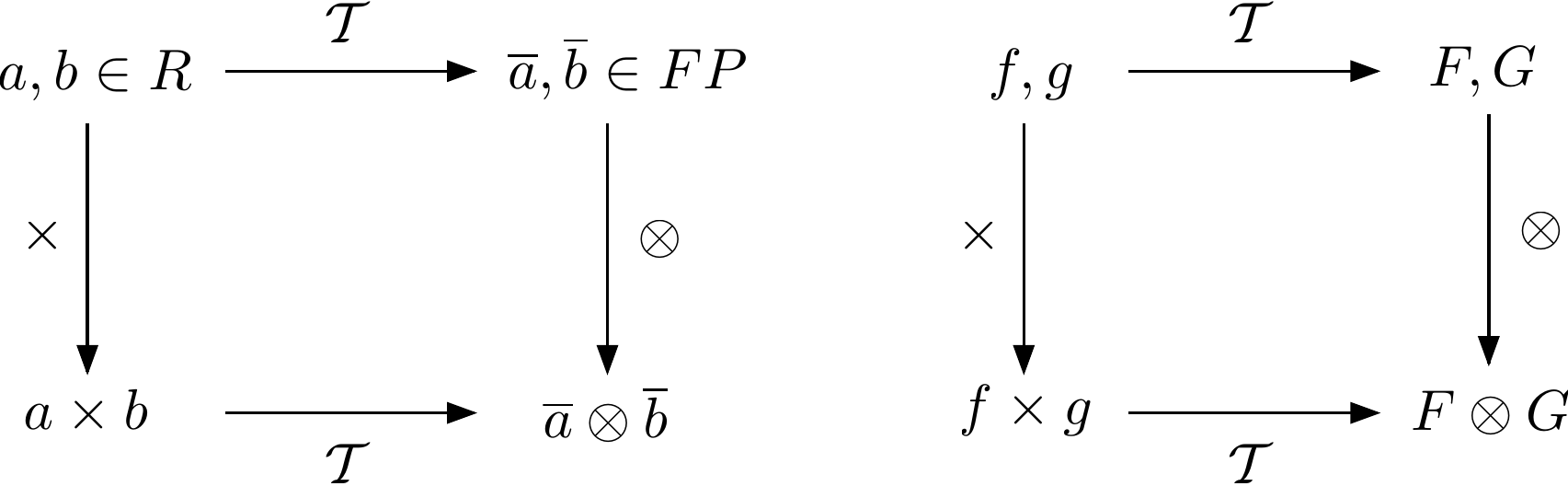}
    \caption{Analogy between the FP representation of real numbers in computer environment (left) and the algebra of Taylor polynomials in DA framework (right)\citep{DiLizia2008a}.}
    \label{fig:DA_scheme}
\end{figure}

Differential algebra provides the tools to compute the derivatives of functions within a computer environment \citep{Ritt1932,Ritt1948, Risch1969, Risch1970,Kolchin1973,Berz1999}. Historically, the treatment of functions in numerics has been based on the treatment of numbers, and the classical numerical algorithms are based on the evaluation of functions at specific points. The basic idea of DA is to bring the treatment of functions and the operations on them to the computer environment in a similar way as the treatment of real numbers \citep{Berz1999}. Real numbers, indeed, are approximated by floating point (FP) numbers with a finite number of digits. With reference to Fig.~\ref{fig:DA_scheme}, let us consider two real numbers $a$ and $b$, and their FP counterpart $\bar{a}$ and $\bar{b}$ respectively: given any operation $\times$ in the set of real numbers, an adjoint operation $\otimes$ is defined in the set of FP numbers so that the diagram in figure commutes. Consequently, transforming the real numbers $a$ and $b$ in their FP representation and operating on them in the set of FP numbers returns the same result as carrying out the operation in the set of real numbers and then transforming the achieved result in its FP representation. In a similar way, suppose two sufficiently regular functions $f$ and $g$ are given. In the framework of DA, these functions are converted into their Taylor series expansions, $F$ and $G$ respectively. In this way, the transformation of real numbers in their FP representation is now substituted by the extraction of the Taylor expansions of $f$ and $g$ (see Fig.~\ref{fig:DA_scheme}, right). For each operation in the function space, an adjoint operation in the space of Taylor polynomials is defined such that the corresponding diagram commutes.

The implementation of DA in a computer environment provides the Taylor coefficients of any function of $v$ variables up to a specific order $n$.
More specifically, by substituting classical real algebra with the implementation of a new algebra of Taylor polynomials, any function $f$ of $v$ variables can be expanded into its Taylor expansion up to an arbitrary order $n$, along with the function evaluation, with a limited amount of effort. The Taylor coefficients of order $n$ for sum and product of functions, as well as scalar products with real numbers, can be directly computed from those of summands and factors. As a consequence, the set of equivalence classes of functions can be endowed with well-defined operations, leading to the so-called truncated power series algebra.
In addition to basic algebraic operations, differentiation and integration can be easily introduced in the algebra, thus finalizing the definition of the differential algebra structure of DA \citep{Berz1986,Berz1987}. The DA used in this work is implemented in the DACE software \citep{Rasotto2016}.

A relevant application of DA is the automatic high order expansion of the solution of an ODE  with respect to the initial conditions \citep{Berz1999,DiLizia2008,Rasotto2016}. This expansion can be achieved by considering that any integration scheme, explicit or implicit, is characterized by a finite number of algebraic operations, involving the evaluation of the ODE right hand side (RHS) at several integration points. Therefore, replacing the operations between real numbers with those on DA numbers, it yields to the $n$th order Taylor expansion of the flow of the ODE, $\phi(t;\delta x_0, t_0)=\mathcal{M}_{\phi}(\delta x_0)$, at each integration time, assuming a perturbed initial condition $x_0+\delta x_0$. Without loss of generality, consider the scalar initial value problem:
\begin{equation}
    \dot{x}(t)=f(t,x),\quad x(t_0)=x_0
	\label{eq:IVP}
\end{equation}
and the associated flow $\phi(t;\delta x_0, t_0)$. For simplicity, consider uncertain initial conditions only. Starting from the $n$th order DA representation of the initial condition, $[x_0]=x_0+\delta x_0$, which is a $(n+1)$-tuple of Taylor coefficients, and performing all the operations in the DA framework, we can propagate the Taylor expansion of the flow in $x_0$ forward in time, up to the final time $t_f$.
Consider, for example, the forward Euler's scheme:
\begin{equation}
    x_i=x_{i-1}+f(x_{i-1})\Delta t
	\label{eq:Euler}
\end{equation}
and replace the initial value with the DA expression $[x_0]=x_0+\delta x_0$. The first time step yields
\begin{equation}
    [x_1]=[x_0]+f([x_0])\Delta t
	\label{eq:Euler_DA}
\end{equation}

If the function $f$ is evaluated in the DA framework, the output of the first step, $[x_1]$, is the $n$th order Taylor expansion of the flow $\phi(t;\delta x_0, t_0)$ in $x_0$ for $t=t_1$. Note that, as a result of the DA evaluation of $f([x_0])$, the $(n+1)$-tuple $[x_1]$ may include several non zero coefficients corresponding to high order terms in $\delta x_0$. The previous procedure can be repeated for the subsequent steps. The result at the final step is the $n$th order Taylor expansion of $\phi(t;\delta x_0, t_0)$ in $x_0$ at the final time $t_f$. Thus, the flow of a dynamical system is approximated, at each time step $t_i$, with its $n$th order Taylor expansion in a fixed amount of effort. Any explicit ODE integration scheme can be rewritten as a DA scheme. For the numerical integrations presented in this paper, a DA version of a $7/8$ Dormand Prince (8th order solution for propagation, 7th order solution for step size control) Runge Kutta scheme is used. 

The main advantage of the DA based approach is that there is no need to write and integrate variational equations to obtain high order expansions of the flow. It is therefore independent on the RHS of the ODE and it is computationally efficient. Unfortunately, DA fails to accurately describe, with a single polynomial map,  the evolution in time of an uncertainty set in case of highly nonlinear dynamics or long term propagation. The approximation error is strictly related to the size of the domain the polynomial is defined in \citep{Wittig2015}. The approximation error between an $n+1$ times differentiable function $f\in C^{n+1}$ and its Taylor expansion $P_f$ of order $n$, without loss of generality taken around the origin, is given by Taylor's theorem:
\begin{equation}
    |f(\delta x)-P_f(\delta x)|\leqslant C\cdot \delta x^{n+1}
	\label{eq:Taylor_error}
\end{equation}
for some constant $C>0$. Consider now the maximum error $e_r$ of $P_f$ on a domain $B_r$ of radius $r>0$ around the expansion point. Considering equation~(\ref{eq:Taylor_error}), we obtain:
\begin{equation}
    |f(\delta x)-P_f(\delta x)|\leqslant C\cdot \delta x^{n+1}\leqslant C\cdot r^{n+1}=e_r
	\label{eq:Taylor_radius}
\end{equation}

If the domain of $P_f$ is reduced from $B_r$ to $B_{r/2}$ of radius $r/2$, the maximum error of $P_f$ over $B_{r/2}$ will decrease by a factor $1/2^{n+1}$:
\begin{equation}
    |f(\delta x)-P_f(\delta x)|\leqslant C\cdot \delta x^{n+1}\leqslant C\cdot \left(\frac{r}{2}\right)^{n+1}=\frac{e_r}{2^{n+1}}
	\label{eq:Taylor_error_0.5}
\end{equation}

\begin{figure}
	\includegraphics[trim=3cm 4cm 3cm 3cm, clip=true, width=\columnwidth]{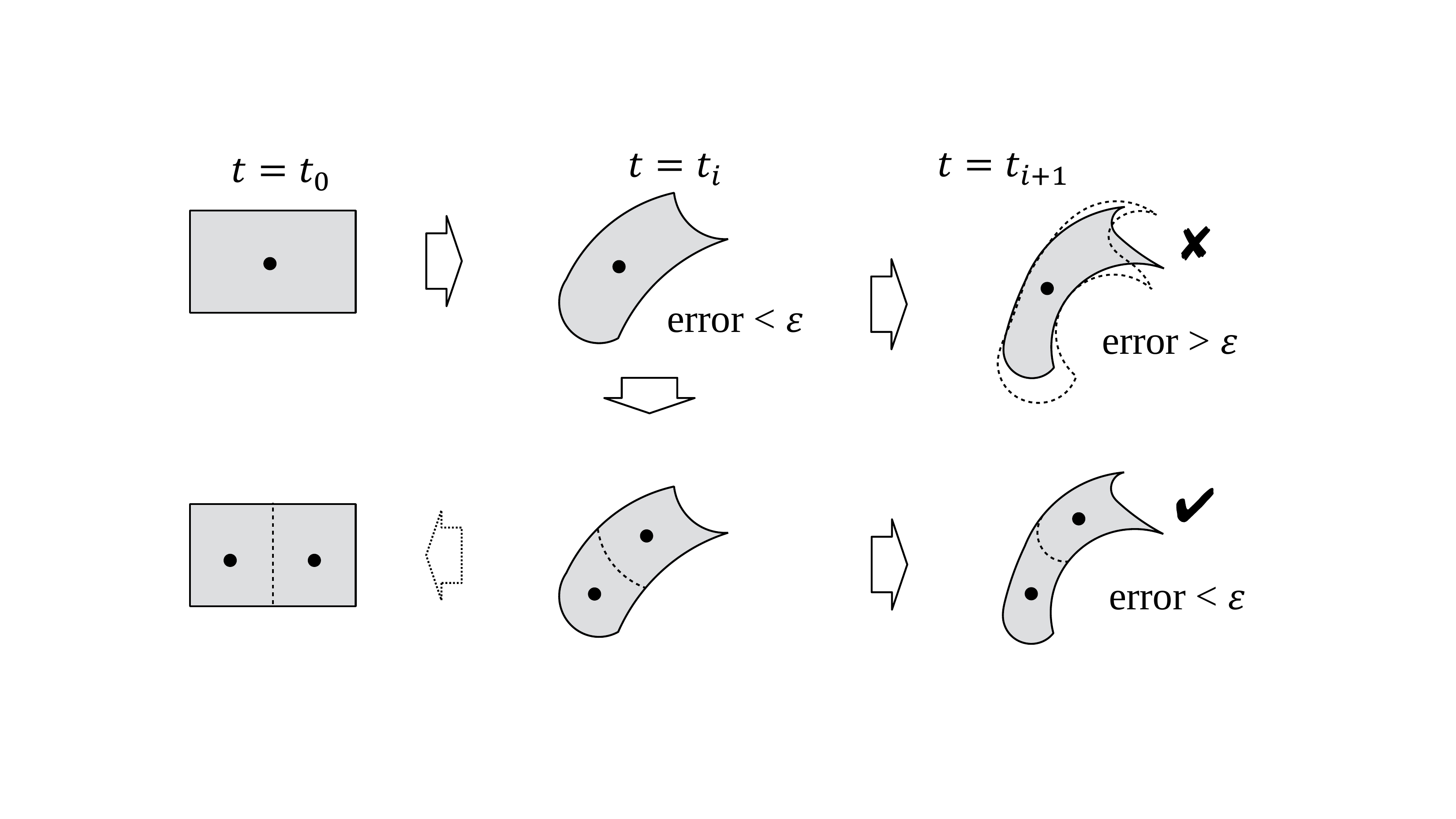}
    \caption{ADS algorithm schematic illustration \citep{Wittig2015}.}
    \label{fig:ADS_scheme}
\end{figure}

By subdividing the initial domain into smaller domains and computing the Taylor expansion around the center points of the new domains, the error greatly reduces, whereas the expansions still cover the entire initial set.
Starting from these considerations, Automatic Domain Splitting (ADS) employs an automatic algorithm to determine at which time $t_i$ the flow expansion over the set of initial conditions is no longer able to describe the dynamics with enough accuracy \citep{Wittig2015}. Once this case has been detected, the domain of the original polynomial expansion is divided along one of the expansion variables into two domains of half their original size. By re-expanding the polynomials around the new centre points, two separate polynomial expansions are obtained. By defining with $x_{i}$ the splitting direction, both generated polynomial expansions $P_1$ and $P_2$ have terms of order $n$ in $x_{i}$ smaller by a factor of $2^n$ with respect to the original polynomial expansion $P$. Thus, the splitting procedure guarantees a more accurate description of the whole uncertainty set at the current time epoch $t_{i}$. After such a split occurs, the integration process is resumed on both generated subsets, until new splits are required.  A representation of the ADS procedure is shown in Fig.~\ref{fig:ADS_scheme}.

The decision on the splitting epoch and, in case of multivariate polynomials, the splitting direction relies on estimating the size of the ($n+1$)th order terms of the polynomial using an exponential fit of the size of all the known non-zero terms up to order $n$. If the size of the truncated order becomes too large, we decide to split the polynomial. This method allows us to consider all the information available in the polynomial expansion and to obtain an accurate estimate of the size of the $n+1$ order term, the first discarded order. The exponential fit is chosen because, after reducing the domain with a sufficient number of splits, the coefficients of the resulting polynomial expansion decay exponentially as a direct consequence of Taylor's theorem. A mathematical description is offered hereafter and follows the scheme presented in \citet{Wittig2015}. Consider a polynomial $P$ of order $n$ of the form
\begin{equation}
    P(\mathbfit{x})=\sum\limits_{\alpha}a_\alpha \mathbfit{x}^\alpha
	\label{eq:Poly}
\end{equation}
written using multi-index notation, the size $S_i$ of the terms of order $i$ is computed as the sum of the absolute values of all coefficients of exact order $i$:
\begin{equation}
    S_i=\sum\limits_{|\alpha|=i}|a_\alpha|
	\label{eq:Poly}
\end{equation}

\begin{figure}
	\includegraphics[trim=0cm 0cm 0cm 0cm, clip=true, width=0.9\columnwidth]{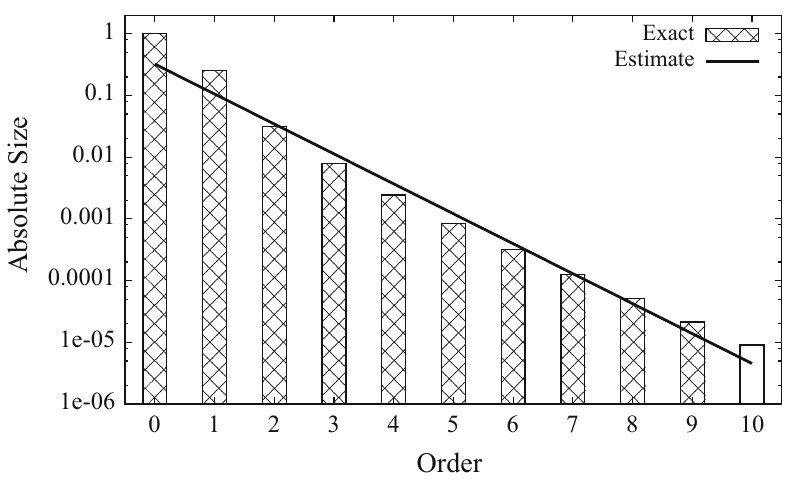}
    \caption{Truncation error estimation for the Taylor expansion of $\sqrt{1+x/2}$ via exponential fitting. Terms of order up to 9 are used for the fitting \citep{Wittig2015}.}
    \label{fig:Err_est}
\end{figure}

We denote by $I$ the set of indices $i$ for which $S_i$ is non-zero. A least squares fit of the exponential function
\begin{equation}
    f(i)=Ae^{Bi}
	\label{eq:Poly}
\end{equation}
is used to determine the coefficients $A$ and $B$ such that $f(i)=S_i,\, i \in I$, is approximated optimally in least squares sense. Then, the value of $f(n+1)$ is used to estimate the size $S_{n+1}$ of the truncated order $n+1$ of $P$. An example of the application of the method is shown in Fig.~\ref{fig:Err_est}, where the polynomial is the Taylor expansion of $\sqrt{1+x/2}$ up to order 9. The size $S_i$ of each order is shown as bars, whereas the resulting fitted function $f$ is shown as a line.

In the case of multivariate polynomials $P(\mathbfit{x})=P(x_1,x_2,\ldots{},x_v)$, the split is performed in one component $x_i$. We determine the splitting direction using a method similar to the one adopted for the splitting decision. For each $j=1,\ldots{},v$ we begin by factoring the known coefficients of $P$ of order up to $n$ with respect to $x_j$, i.e.
\begin{equation}
    P(x_1,x_2,\ldots{},x_v)=\sum\limits_{m=0}^{n} x_j^m\cdot q_{j,m}(x_1,\ldots{},x_{j-1},x_{j+1},\ldots{},x_v)
	\label{eq:Poly_factor}
\end{equation}
where the polynomials $q_{j,m}$ do not depend on $x_j$. The size $S_{j,m}$ of the polynomials $q_{j,m}$ is estimated by the sum of the absolute values of their coefficients. Then, the exponential fitting routine is applied to estimate the size $S_{j,n+1}$ of the truncated terms of order $n+1$ in $x_j$. Finally, the splitting direction $i$ is chosen as the component $x_j$ with the largest truncation error $S_{j,n+1}$. In this way, all splits are performed in the direction of the variable that currently has the largest estimated contribution to the total truncation error of the polynomial $P$.

The main parameters of the algorithm are the tolerance for the splitting procedure and the maximum number of allowed splits $N_{\text{max}}$. The first parameter is selected according to the required precision of the polynomial expansions and determines the splitting epochs: when the estimated truncation error exceeds the imposed tolerance, the current domain is split. As a direct consequence of the ADS procedure, the maximum error over the obtained set of polynomials decreases with the selected splitting precision. However, the maximum error is always larger than the selected integration precision. This difference is actually expected, as the splitting tolerance plays a similar role as the one-step error set in the automatic step size control of the integration scheme \citep{Wittig2015}. It is the maximum error that can accumulate at any time before the integrator takes action to reduce further error accumulation. However, the accumulated error at the time of the splitting cannot be undone as the splitting only re-expands the polynomial to prevent exponential growth in future integration steps. The ideal tolerance depends on both the dynamics and the integration time, and it has to be chosen heuristically to ensure the final result satisfies the accuracy requirements of the application. A numerical example is shown in Section~\ref{Sub_eps}. 

The second parameter plays the role of limiting the number of generated subdomains by imposing a minimum size for the generated subsets: domain splitting is disabled on any set whose volume is less than $2^{-N_{\text{max}}}$ times that of the initial domain. That is, any set is split at most $N_{\text{max}}$ times. Then, instead of splitting a set further, integration is stopped at the attempt to perform the $(N_{\text{max}}+1)$th split and the resulting polynomial expansion is saved as incomplete. Incomplete polynomials are later treated separately in the analysis of the results \citep{Wittig2015}. 

When each generated subset reaches either the final simulation time or the minimum box size, the ADS propagation terminates, and the result is a list of polynomial expansions, each covering a specific subset of the domain of initial conditions. A more detailed description of the ADS algorithm can be found in \citet{Wittig2015}. 

\section{Automatic Domain Pruning}
\label{Section_ADP}
As described in Section~\ref{Section_ADS}, automatic domain splitting provides an accurate description of the evolution in time of a given uncertainty set by splitting the domain in subsets when required. Unfortunately, this method may entail a not negligible computation effort, as all generated subsets are propagated to the final simulation time or till the minimum box size is reached. While this approach is unavoidable when the behaviour of the whole uncertainty set is analysed, it becomes a strong limitation when only a portion of the initial set is to be investigated. This is the case when the first resonant return of a Near Earth Object is studied. Resonant returns occur when, during a close encounter, an asteroid is perturbed into an orbit with a period $T'\sim k/h$ years. Thus, after $h$ revolutions of the asteroid and $k$ revolutions of the Earth, both celestial bodies are in the same region of the first close encounter and a second one may occur. Given the initial uncertainty set, only a portion of it may lead to the resonant return. It would be therefore interesting to limit the propagation to this region only.

Starting from these considerations, the Automatic Domain Pruning (ADP) we present in this paper combines the ADS algorithm with a pruning technique with the aim of limiting the number of propagated subsets. We make here the assumption of no close approaches with other celestial bodies between the first close encounter and the selected resonant return. This assumption is easily checked right before the ADP propagation, as later explained in Section~\ref{Sub_DT_ref}. 

The first phase of the algorithm consists in propagating the whole uncertainty set by means of ADS propagation up to the epoch of the first close encounter. The availability of the polynomial expansion of the state vector of the object with respect to the initial uncertainty provides the polynomial expansion of the orbital period of the object after the close encounter. By using a polynomial bounder, we can estimate the range of all possible values of the orbital period after the close encounter and, thus, retrieve all possible resonances with the planet, i.e. all orbital periods included in the computed orbital period range leading to a resonant return with the planet. 

\begin{figure}
	\includegraphics[trim=1cm 0cm 0.5cm 0cm, clip=true,width=\columnwidth]{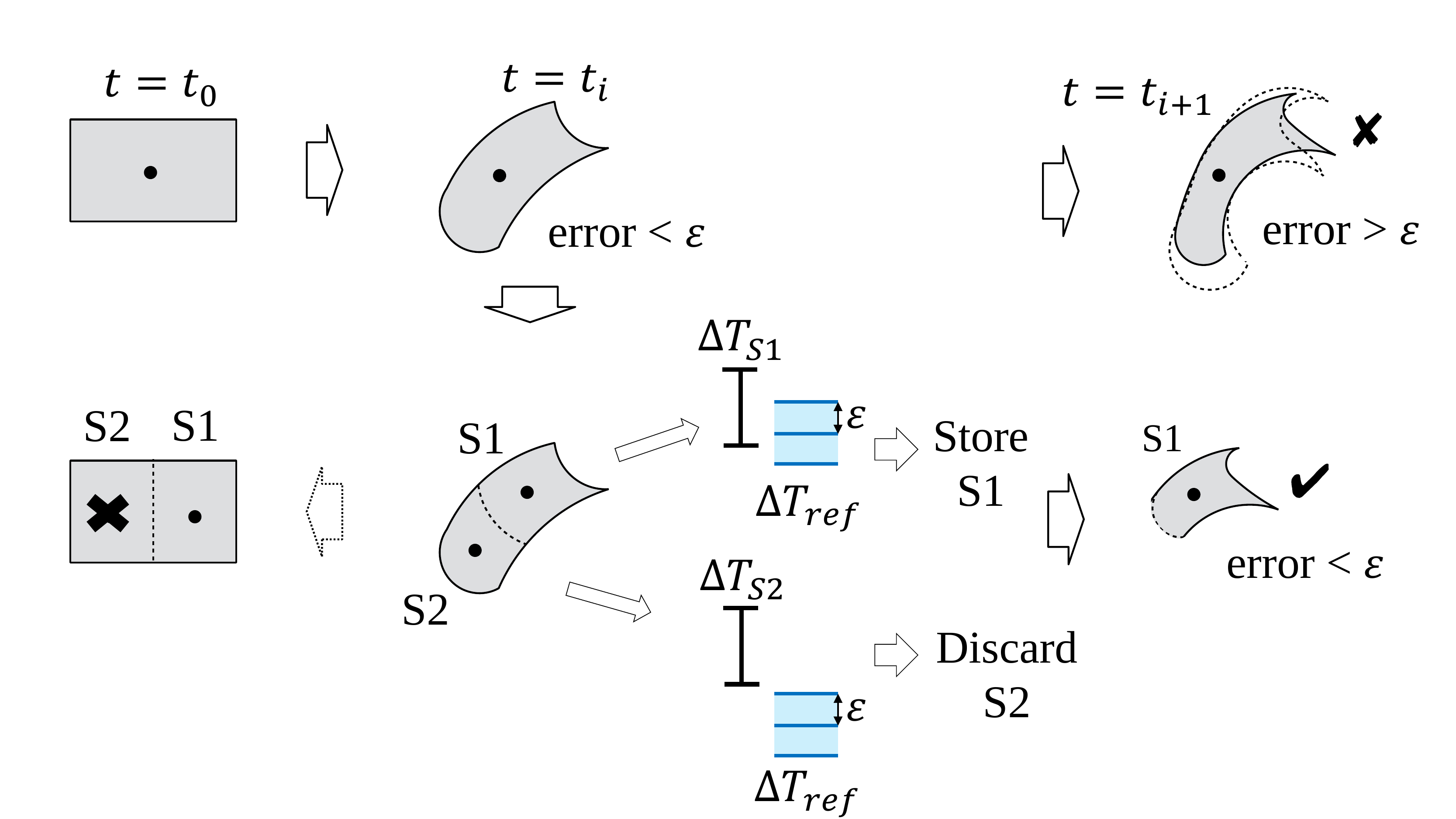}
    \caption{ADP algorithm illustration. Pruning is performed by comparing the estimated subset orbital period range $\Delta T_{\text{sub}}$ with the reference range $\Delta T_{\text{ref}}$.}
    \label{fig:ADP_scheme}
\end{figure}

Once all resonances are computed, the analysis focuses on a single resonance, and the propagation is resumed. Every time a new subset is generated, the method automatically identifies if the set may lead to the investigated resonant return or not. By exploiting the knowledge of the DA state vector at the epoch of the first close encounter, indeed, we can assign a given orbital period range to each generated subset. This range, defined as $\Delta T_{\text{sub}}$, is compared to a reference range $\Delta T_{\text{ref}}$, centred in the resonance period $T'$ with a semi-amplitude $\varepsilon$, $\Delta T_{\text{ref}}=[(1-\varepsilon)T', (1+\varepsilon)T']$. We select the reference range in order to consider small dynamical perturbations between the first close encounter and the resonant return. If $\Delta T_{\text{sub}}$ is at least partially included in the reference range, then the current subset is retained, and its propagation is continued. If $\Delta T_{\text{sub}}$ is not included in the reference range, then the initial conditions included in the current subset do not lead to a resonant return at the investigated epoch, and so the subset is discarded. This way, subsets are dynamically pruned during the ADS propagation. An illustration of the ADP algorithm is shown in Fig.~\ref{fig:ADP_scheme}.

The ADP algorithm, therefore, does not alter the sequence of generated subdomains, but limits the propagation in time to those subsets that are involved in the investigated resonant return. This pruning action has a positive impact on the overall computational burden, since the computational effort required by the propagation of all the discarded subsets is saved. As only subsets with close approaches to the Earth at the epoch of the investigated resonant return are maintained, the result at the end is a set of subdomains whose propagation stops slightly before the epoch of the investigated resonant return for having reached their minimum box size.

\section{Importance Sampling Method}
\label{Section_IS}
The output of the ADP propagation is a list of subsets at epochs close to the investigated resonant return. Still, no value for the impact probability is available. We obtain an estimate for the impact probability by sampling the generated subsets and propagating the samples till they reach their minimum geocentric distance. Among all possible sampling technique, we employ the Importance Sampling (IS) method \citep{Zio2013}. 

The IS method amounts to replacing the original probability density function (pdf) $q_{\mathbfit{x}}(\mathbfit{x})$ with an Importance Sampling Distribution (ISD) $\tilde{q}_{\mathbfit{x}}(\mathbfit{x})$ arbitrarily chosen by the analyst so as to generate a large number of samples in the importance region of the phase space $F$, the region of initial conditions leading to an impact with Earth at the epoch of the resonant return. In the case under study, we select the auxiliary distribution in order to limit as much as possible the generation of the samples to the subsets that get through the dynamic pruning. The IS algorithm is the following:

\begin{enumerate}
\item Identify a proper $\tilde{q}_{\mathbfit{x}}(\mathbfit{x})$.

\item Express the impact probability $p(F)$ as a function of $\tilde{q}_{\mathbfit{x}}(\mathbfit{x})$.
\begin{equation}
    p(F)=\int I_{\text{F}}(\mathbfit{x})q_{\mathbfit{x}}(\mathbfit{x})\mathrm{d}\mathbfit{x}=\int \frac{I_{\text{F}}(\mathbfit{x})q_{\mathbfit{x}}(\mathbfit{x})}{\tilde{q}_{\mathbfit{x}}(\mathbfit{x})}\tilde{q}_{\mathbfit{x}}(\mathbfit{x})\mathrm{d}\mathbfit{x}
	\label{eq:IS_integral}
\end{equation}
where $I_{\text{F}}(\mathbfit{x}):{\rm I\!R}^{v}\rightarrow {\{0,1\}}$ is an indicator function such that $I_{\text{F}}(\mathbfit{x})=1$ if $\mathbfit{x}\in F$, $0$ otherwise.

\item 	Draw $N_{\text{T}}$ samples ${\mathbfit{x}^k:k=1, 2, \ldots{},N_{\text{T}}}$ from the importance sampling distribution $\tilde{q}_{\mathbfit{x}}(\mathbfit{x})$. If a good choice for the auxiliary pdf is made, the generated samples concentrate in the region $F$.

\item 	Compute the estimate $\hat{p}(F)$ for the impact probability $p(F)$ by resorting to equation~(\ref{eq:IS_integral}):
\begin{equation}
    \hat{p}(F)=\frac{1}{N_{\text{T}}}\sum_{k=1}^{N_{\text{T}}} \frac{I_{\text{F}}(\mathbfit{x}^k)q_{\mathbfit{x}}(\mathbfit{x}^k)}{\tilde{q}_{\mathbfit{x}}(\mathbfit{x}^k)}
	\label{eq:IS_summation}
\end{equation}

\item Compute the variance of the estimator $\hat{p}(F)$ as:
\begin{equation}
\begin{aligned}
    \sigma^2(\hat{p})=\frac{1}{N_{\text{T}}}\left(\int \frac{I_{\text{F}}^2(\mathbfit{x})q_{\mathbfit{x}}^2(\mathbfit{x})}{\tilde{q}_{\mathbfit{x}}^2(\mathbfit{x})}\tilde{q}_{\mathbfit{x}}(\mathbfit{x})\mathrm{d}\mathbfit{x}-p^2(F) \right)\\
    \approx \frac{1}{N_{\text{T}}}\left(\widehat{p^2(F)}-\hat{p}^2(F)\right)
	\label{eq:IS_variance}
\end{aligned}
\end{equation}

\end{enumerate}

The selection of the ISD represents the most critical point for the method. Several techniques have been developed in order to find the one giving small variance for the estimator \citep{Zio2013}. In this paper, we shape the ISD according to the result of the ADP propagation. As described in Section~\ref{Section_ADP}, the ADP propagation provides a list of subsets whose propagation is stopped slightly before the resonant return. All subsets are identified as Potentially Hazardous Subdomains (PHS's), but no probability ranking is provided by the ADP propagation. Starting from these considerations, we define the ISD as a uniform probability density function including all the generated subsets over the whole domain. This selection allows us to increase the number of samples drawn in the PHS's and, eventually, in the impact-leading region.

\section{Automatic Domain Pruning Importance Sampling Method}
\label{Section_ADP-IS}
The combination of the methods presented in Sections~\ref{Section_ADP} and ~\ref{Section_IS} yields the ADP importance sampling method (ADP--IS) for uncertainty propagation and impact probability computation of the first resonant returns of NEO. The starting point is represented by the output of an orbit determination process of a given NEO at the observation epoch $t_0$. This output can be expressed in terms of estimated state vector and related covariance matrix. Then, the steps of the ADP propagation phase are the following:
\begin{enumerate}
\item 	Consider the initial state vector and related pdf and perform an analysis to identify possible epochs of close encounters and resonant returns. The analysis is carried out by propagating the uncertainty set using ADS up to the first close encounter, computing the semi-major axis dispersion over the set with a polynomial bounder and identifying the resonant frequencies. The validity of the resonances is then checked as explained in Section~\ref{Sub_DT_ref}.

\item Select a resonance and identify its epoch $t_{\text{res}}$.

\item Perform an ADP propagation till the epoch $t_{\text{res}}$. Every time a split is required, compare the orbital period range of the current subset $\Delta T_{\text{sub}}$ with the reference range $\Delta T_{\text{ref}}$: 
\begin{equation}
    \Delta T_{\text{sub}} \cap \Delta T_{\text{ref}} \begin{cases}
    =0 \quad \text{discard\,the\,current\, subset}\\
    \neq 0 \quad \text{include\,the\,current\,subset}
\end{cases}
	\label{eq:ADP_check}
\end{equation}

\end{enumerate}

The method provides a set of $n_{\text{PHS}}$ PHS's and related DA state vectors $[\mathbfit{x}_f^{i}]$ at the truncation time $t_{f}^{i}$, with $i=1,\ldots{}, n_{\text{PHS}}$. Vector $[\mathbfit{x}_{f}^{i}]$ is a polynomial state vector, each component being a function of the initial conditions $\mathbfit{x}_0^{i}$.

The IS phase is initialized by setting the value of the estimated impact probability $\hat{p}_{\text{old}}$ and the number of iterations $n_{it}$  equal to zero. Then, the steps of the algorithm are the following:
\begin{enumerate}
\item 	Define the ISD function $\tilde{q}_{\mathbfit{x}}(\mathbfit{x})$ as a uniform pdf including all the generated PHS's.

\item \label{IS.2} Set $n_{it}=n_{it}+1$ and draw one sample $\mathbfit{x}_0^{it}$ from $\tilde{q}_{\mathbfit{x}}(\mathbfit{x})$.

\item Check if the sample belongs to one of the PHS's: if it is out of the PHS's, go back to step~\ref{IS.2}, otherwise identify the correct PHS $i$ the sample belongs to.

\item 	Compute the algebraic state vector $\mathbfit{x}_{f}^{it}$ corresponding to the drawn sample $\mathbfit{x}_0^{it}$ at the truncation epoch $t_{f}^{i}$ by performing a polynomial evaluation of the DA state vector $[\mathbfit{x}_{f}^{i}]$ at $\mathbfit{x}_0^{it}$.  That is, $\mathbfit{x}_{f}^{it}=[\mathbfit{x}_{f}^{i}](\mathbfit{x}_0^{it})$.

\item 	Propagate the state vector $\mathbfit{x}_{f}^{it}$ from $t_{f}^{i}$ to the epoch of the selected resonant return.

\item 	Compute the minimum geocentric distance $d|_{\text{res}}^{it}$ and evaluate the indicator $I_{\text{F}}^{it}$
\begin{equation}
    I_{\text{F}}^{it}= \begin{cases}
    0 \quad \text{if}\,d|_{\text{res}}^{it}>R_{\earth}
    \\
    1 \quad \text{if}\,d|_{\text{res}}^{it}<R_{\earth}
\end{cases}
	\label{eq:Indicator}
\end{equation}

\item 	If $I_{\text{F}}^{it}=0$, go back to step~\ref{IS.2}, otherwise evaluate the new impact probability $\hat{p}_{\text{new}}$. By reformulating equation~(\ref{eq:IS_summation}), we obtain:
\begin{equation}
   \hat{p}_{\text{new}}=\frac{1}{N_{\text{T}}}\sum_{k=1}^{N_{\text{T}}} \frac{I_{\text{F}}(\mathbfit{x}^{k})q_{\mathbfit{x}}(\mathbfit{x}^{k})}{\tilde{q}_{\mathbfit{x}}(\mathbfit{x}^{k})}=\frac{1} {n_{it}}(\hat{I}+q_{\mathbfit{x}}(\mathbfit{x}_0^{it}))\frac{1}{\tilde{q}_{\mathbfit{x}}(\mathbfit{x}_0^{it})}
	\label{eq:p_new}
\end{equation}
where $q_{\mathbfit{x}}(\mathbfit{x}_0^{it})$ is the value of the original pdf in $\mathbfit{x}_0^{it}$, $\tilde{q}_{\mathbfit{x}}(\mathbfit{x}_0^{it})$ is the value of the auxiliary pdf in $\mathbfit{x}_0^{it}$, whereas the term $\hat{I}$ represents the summation of all terms $I_{\text{F}}(\mathbfit{x}_0^{k})q_{\mathbfit{x}}(\mathbfit{x}_0^{k})$ of the previous iterations. The total number of samples considered for the estimation is $n_{it}$, i.e. the number of drawn samples when the estimate is computed. Note that, since the ISD is uniform over the whole set of PHS's, it can be extracted from the summation.

\item 	Compare $\hat{p}_{\text{old}}$ and $\hat{p}_{\text{new}}$: if the relative difference is larger than an imposed tolerance, go back to step ~\ref{IS.2}, otherwise stop.

\end{enumerate}

\section{Numerical simulations: the case of asteroid (99942) Apophis}
\label{Section_Apophis}
In this section, we assess the performance of the ADP--IS method on the evaluation of the impact probability for the test case of asteroid (99942) Apophis. Table~\ref{tab:Apophis_IC} shows the nominal initial state and associated uncertainties $\sigma$ for Apophis on June 18, 2009 expressed in terms of equinoctial parameters $\mathbfit{p}=(a,P_1,P_2,Q_1,Q_2,l)$, considering a diagonal covariance matrix. Data were obtained from the Near Earth Objects Dynamic Site\footnote{\url{http://newton.dm.unipi.it/neodys/}} in September 2009.

We selected a diagonal covariance matrix in order to help distinguish the contribution of the six orbital parameters and test our method in a scenario in which the uncertainty volume is maximized. In general, however, this selection may lead to quite inaccurate results as uncertainties may be highly correlated. Nevertheless,the method can be applied in the most general case of full covariance matrix exactly in the same way, with the only difference that the DA variables would be placed along the directions of the covariance eigenvectors to avoid artificially adding extra-volume in the initial domain definition.

As previously stated, the starting point, not including recent optical and radar observations performed from late 2011 onward, was selected in order to test the algorithm against the most critical scenario. Asteroid Apophis will have a close encounter with Earth on April 13, 2029 with a nominal distance of $3.8\cdot 10^4$~km \citep{Chesley2005}. According to the selected initial conditions, though an impact in 2029 can be ruled out, the perturbations induced by the encounter open the door to resonant returns in 2036 and 2037. The aim is therefore to apply the presented method to provide an estimate for the impact probability at the epoch of the first resonant return, in 2036. 

The motion of Apophis in the Solar system is modelled according to the $(N+1)$body problem, including relativistic corrections to the Newtonian forces \citep{Seidelmann1992,Wittig2015}. Specifically, the full equation is
\begin{equation}
\begin{aligned}
   \ddot{\mathbfit{r}}=G\sum\limits_{i}\frac{m_i(\mathbfit{r}_i-\mathbfit{r})}{r_i^3}
   \Bigg\{1-\frac{2(\beta+\gamma)}{c^2}G\sum\limits_j\frac{m_j}{r_j}-\frac{2\beta-1}{c^2}G\sum\limits_{j\neq i}\frac{m_j}{r_{ij}}+\\   
   \frac{\gamma|\dot{\mathbfit{r}}|^2}{c^2}+\frac{(1+\gamma)|\dot{\mathbfit{r}}_i|^2}{c^2}-
   \frac{2(1+\gamma)}{c^2}\dot{\mathbfit{r}}\cdot\dot{\mathbfit{r}}_i-\frac{3}{2c^2}\left[\frac{(\mathbfit{r}-\mathbfit{r}_i)\cdot\dot{\mathbfit{r}}_i}{r_i}\right]^2+\\
   \frac{1}{2c^2}(\mathbfit{r}_i-\mathbfit{r})\cdot\ddot{\mathbfit{r}}_i\Bigg\}+
   G\sum\limits_i\frac{m_i}{c^2r_i}\Bigg\{\frac{3+4\gamma}{2}\ddot{\mathbfit{r}}_i+
   \frac{\{[\mathbfit{r}-\mathbfit{r}_i]\cdot[(2+2\gamma)\dot{\mathbfit{r}}}{r_i^2}-\\\frac{(1+2\gamma)\dot{\mathbfit{r}}_i]\}(\dot{\mathbfit{r}}-\dot{\mathbfit{r}}_i)}{r_i^2}\Bigg\}
	\label{eq:p_new}
\end{aligned}
\end{equation}
where $\mathbfit{r}$ is the position of Apophis in Solar System barycentric coordinates, $G$ is the gravitational constant, $m_i$ and $\mathbfit{r}_i$ are the mass and the Solar System barycentric position of Solar System body $i$, $r_i=|\mathbfit{r}_i-\mathbfit{r}|$, $c$ is the speed of light in vacuum, and $\beta$ and $\gamma$ are the parametrized post-Newtonian parameters measuring the nonlinearity in superposition of gravity and space curvature produced by unit rest mass \citep{Seidelmann1992}.
The position and velocity vectors of all celestial bodies are computed with NASA's SPICE library\footnote{\url{http://naif.jpl.nasa.gov/naif/toolkit.html}}. We used the planetary and lunar ephemeris DE432s. The $N$ bodies include the Sun, the planets and the Moon. For planets with moons, with the exception of the Earth, the centre of mass of the system is considered. The dynamical model is written in the J2000 ecliptic reference frame.

\begin{figure*}
	\includegraphics[width=0.8\textwidth]{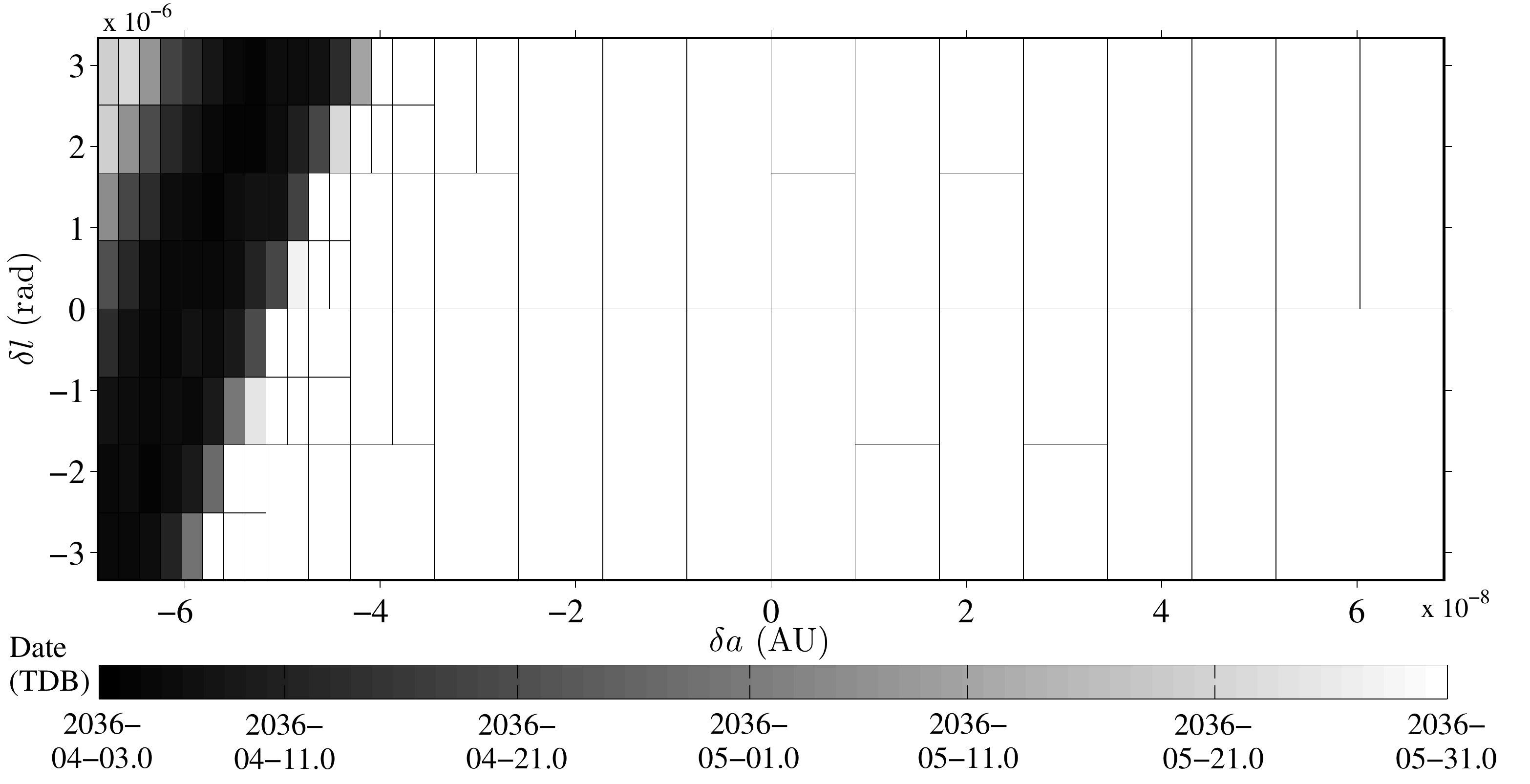}
    \caption{Projection of the generated subsets onto the $a-l$ plane of the initial conditions (ADS propagation, order 8, tolerance $10^{-10}$, $N_{\text{max}}$ 12, $3\sigma$ domain).}
    \label{fig:ADS_result}
\end{figure*}

\begin{table}
	\centering
	\caption{Apophis equinoctial parameters and related uncertainties on June 18, 2009 00:00:00 (TDB).}
	\label{tab:Apophis_IC}
	\begin{tabular}{lccr} 
		\hline
		 & Nominal value & $\sigma$ & \\
		\hline
		$a$ & 0.922438242375914 & $2.29775\cdot 10^{-8}$ & AU\\
		$P_1$ & -0.093144699837425 & $3.26033\cdot 10^{-8}$ & -\\
		$P_2$ & 0.166982492089134 & $7.05132\cdot 10^{-8}$ & -\\
		$Q_1$ & -0.012032857685451 & $5.39528\cdot 10^{-8}$ & -\\
		$Q_2$ & -0.026474053361345 & $1.83533\cdot 10^{-8}$ & -\\
		$l$ & 88.3150906433494 & $6.39035\cdot 10^{-5}$ & $^{\circ}$\\
		\hline
	\end{tabular}
\end{table}

\begin{figure}
	\includegraphics[trim=0cm 0cm 0cm 0cm, clip=true, width=\columnwidth]{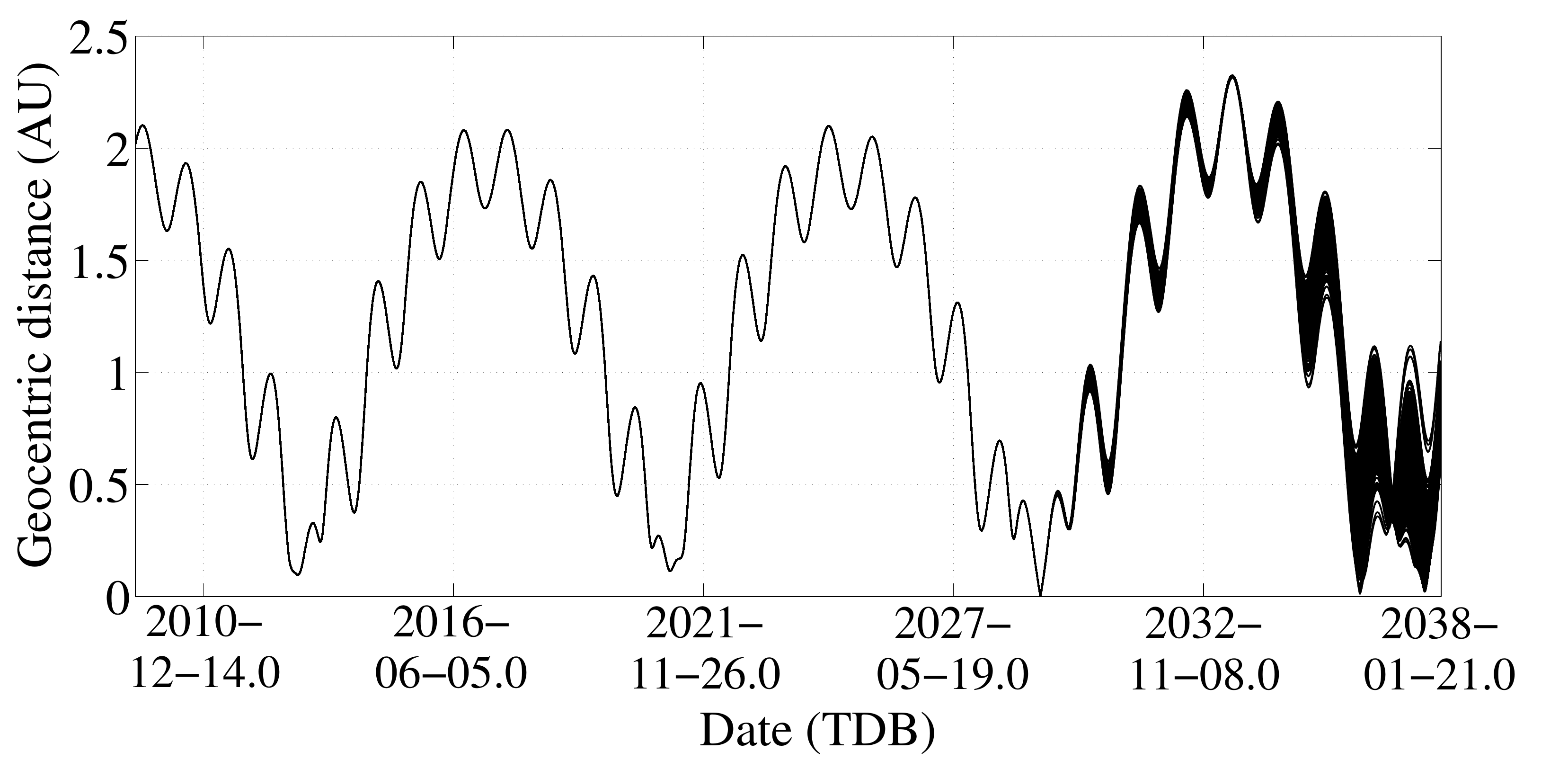}
    \caption{Geocentric distance profile up to January 21, 2038 for one thousand samples from the initial uncertainty set.}
    \label{fig:EarthDistance}
\end{figure}

Figure~\ref{fig:EarthDistance} shows the geocentric distance profile in time for one thousand samples from the initial Gaussian distribution. As expected, the uncertainties significantly increase after 2029 and pave the way to resonant returns in 2036 and 2037.

The authors showed an analysis of the performance of the ADS algorithm for the propagation of the whole set up to the second resonant return in \citet{Wittig2015}. The results are now limited to the first resonant return, and they will be used as a reference for the assessment of the performance of the ADP. 

All the results presented in this section are obtained considering an expansion order equal to 8, a tolerance for the splitting procedure equal to $10^{-10}$, a value of $N_{\text{max}}$ equal to 12 and an initial uncertainty set with $3\sigma$ boundaries, i.e. a 6-dimensional (6D) rectangle with $3\sigma$ boundaries. 

The initial uncertainty set should be properly selected, as the neglected part of the probability mass, i.e. the integral of the pdf over the domain outside the considered box, could significantly alter the estimated impact probability. For the case under study, in which we are considering a 6-dimensional problem with uncorrelated variables, the selection of a 6D rectangular domain with $3\sigma$ boundaries corresponds to considering the 98.4 per cent of the probability mass, and so the estimated impact probability may result underestimated. The accuracy of the estimate improves for larger initial uncertainty sets. A detailed sensitivity analysis on the uncertainty set size and all the other available parameters is offered in Section~\ref{Sensitivity}. All computations are performed on a single core Intel i7-3770 CPU @3.4 GHz, 16 GB RAM processor. 

\begin{figure*}
	\includegraphics[trim=0cm 0cm 0cm 0cm, clip=true,width=0.8\textwidth]{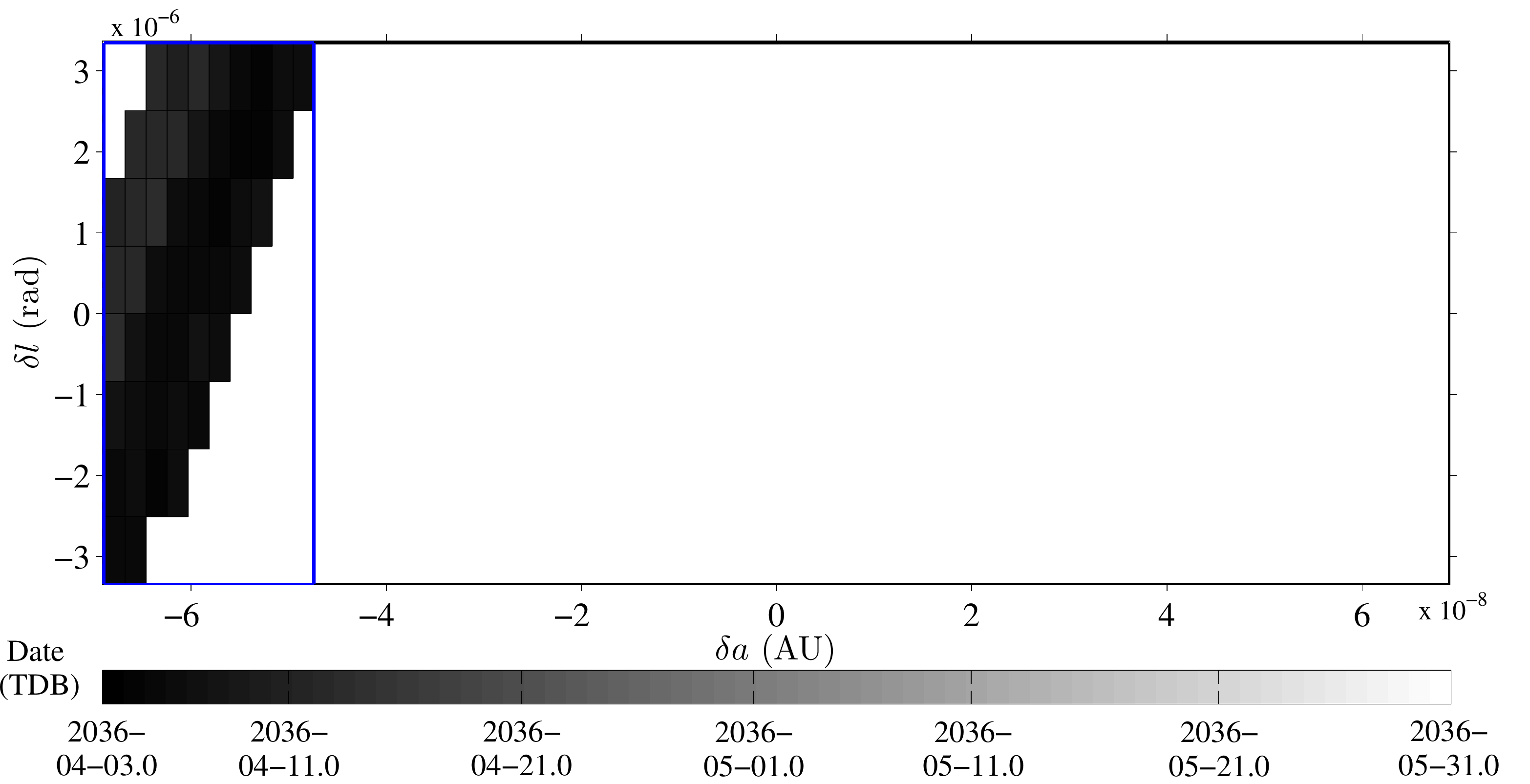}
    \caption{Projection of the generated subsets onto the $a-l$ plane of the initial conditions (ADP propagation, order 8, tolerance $10^{-10}$, $N_{\text{max}}$ 12, $3\sigma$ domain). In blue, boundaries of the ISD.}
    \label{fig:ADP_result}
\end{figure*}

\begin{figure}
	\includegraphics[trim=0cm 0cm 0cm 0cm, clip=true,width=\columnwidth]{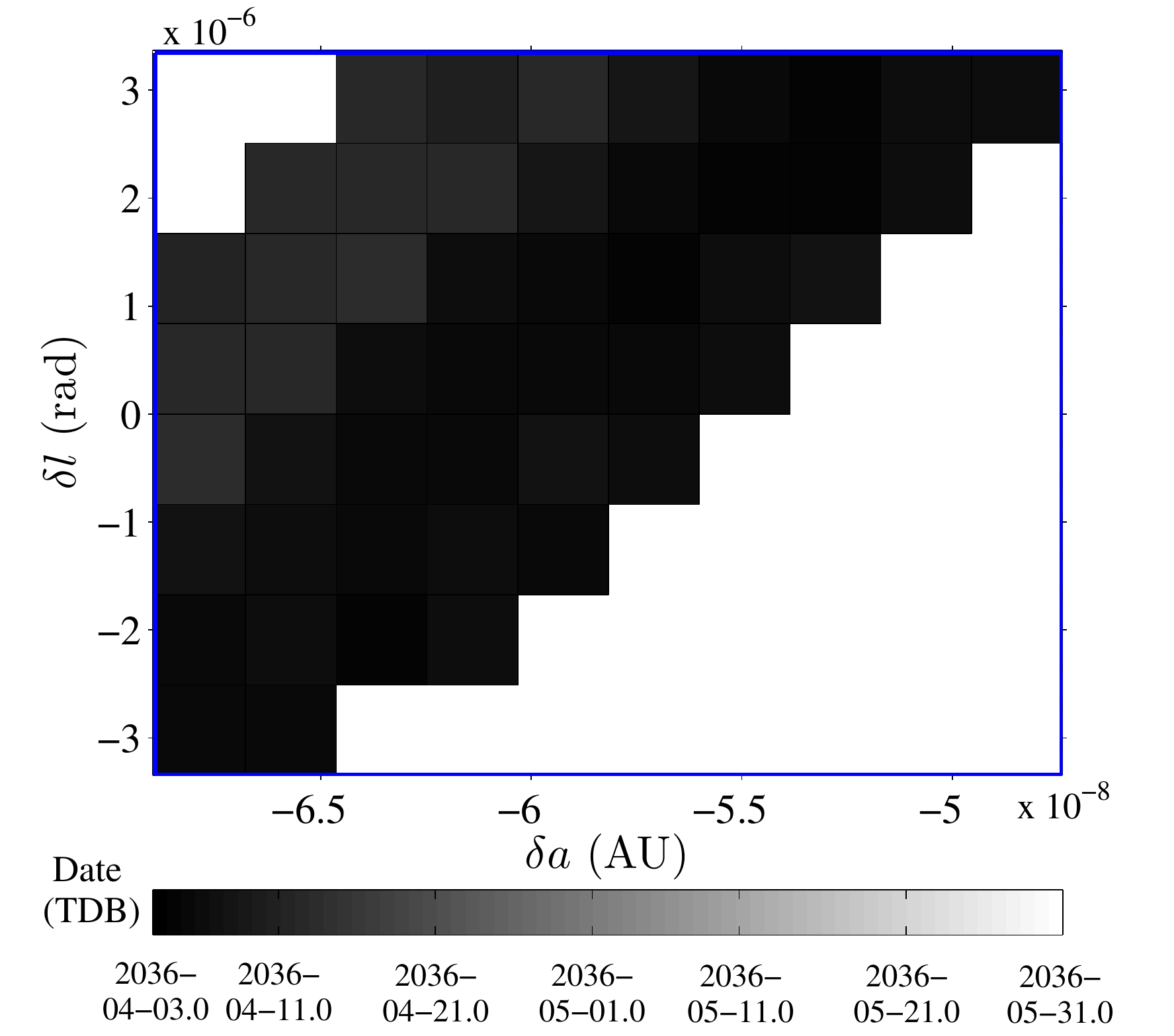}
    \caption{Projection of the generated subdomains onto the $a-l$ plane (detail of Fig.~\ref{fig:ADP_result}).}
    \label{fig:ADP_result_focus}
\end{figure}

\begin{figure}
	\includegraphics[width=\columnwidth]{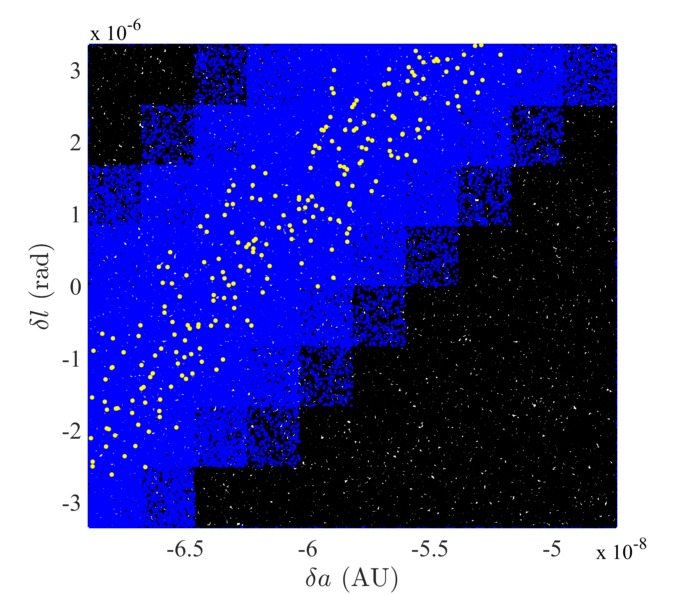}
    \caption{Projection of the generated samples onto the $a-l$ plane. In black, discarded samples. In blue, samples belonging to the PHS's. In yellow, impacting samples.}
    \label{fig:ADP-IS_result}
\end{figure}

The number of subdomains obtained with ADS propagation without pruning is 653, while the computational time is 10~h~6~min. An analysis of the average number of splits per direction shows that most splits occur in the semi major axis ($a$) and true longitude ($l$) directions \citep{Wittig2015}. Thus, though the problem is six dimensional, the analysis on the dynamics can be focused on the projection onto the $a-l$ plane of the initial conditions. 

Figure~\ref{fig:ADS_result} shows the projection of the initial uncertainty box onto the $a-l$ plane, along with the subdomains generated during the ADS propagation. Colours refer to the truncation epoch of the related subset: white regions represent subsets that were able to reach the final simulation time (May 31, 2036, after the expected resonant return), coloured regions represent subsets whose propagation was stopped earlier because they reached their minimum box size. Figure~\ref{fig:ADS_result} can be exploited to easily identify the regions of the initial set that are involved in the resonant return in 2036. While all initial conditions lying within white regions have no risk to impact the Earth, coloured subdomains represent sets of initial conditions that might lead to close encounters with Earth at that epoch. That is, coloured regions represent PHS's. This behaviour is expected, as splits occur when the nonlinearities increase, which happens when trajectories get closer to Earth. It is evident, however, that a significant portion of the computational effort required by the ADS propagation is spent on regions of the initial set that are not involved in the first resonant return. Thus, the application of a selective pruning technique as the ADP aims at alleviating this inefficiency. 

We now investigate the performance of the ADP method. The first part of the analysis is represented by the propagation of the uncertainty set up to the epoch of the first close encounter in 2029. The DA propagation of the whole uncertainty set up to the close encounter in 2029 is performed with no splits. Therefore, the whole set can be described with a single polynomial map at the epoch of the first close encounter. The availability of the DA state vector of the asteroid, then, provides the polynomial expansion of its perturbed orbital period immediately after the close encounter with the Earth. This polynomial expansion allows us to estimate the asteroid orbital period range after the close encounter by means of a polynomial bounder: for the case under study, this range is equal to [415.02, 428.91]~days. By looking at this range, we can identify the first resonances: $T_{\text{res1}}=7/6 T_{\earth}=426.12$~days (where $T_{\earth}$ is the Earth orbital period) is the first resonant orbital period included in the computed range, and it represents a resonant return in 2036. This value is expected, as shown in Fig.~\ref{fig:EarthDistance}. We can also notice that the expected second resonant return (in 2037, resonance 8:7), is also included ($T_{\text{res2}}=417.43$ days).

The a priori identification of the resonances and the application of the ADP--IS method is strictly related to the assumption of no intervening close encounter with other major bodies in between. This assumption is checked immediately after the resonances computation, as explained in Section~\ref{Sub_DT_ref}. For the case under study, the assumptions are verified. Therefore, we can now concentrate the analysis on the first resonant return, in 2036. Given a nominal value $T'=7/6 T_{\earth}$, $\Delta T_{\text{ref}}$ is determined by setting a value of $\varepsilon$ equal to $10^{-3}$. The value of $\varepsilon$ is selected in order to take into account small perturbations between the close encounter in 2029 and the resonant return in 2036. An analysis of the impact of $\varepsilon$ on the results is carried out in Section~\ref{Sub_DT_ref}. The propagation is then resumed as described in Section~\ref{Section_ADP-IS}. Figure~\ref{fig:ADP_result} shows the results of the ADP propagation in terms of subdomains distribution on the $a-l$ plane. A comparison with Fig.~\ref{fig:ADS_result} clearly shows how the ADP restricts the propagation of the generated subdomains to a limited portion of the PHS's. That is, only subsets that are actually involved in the resonant return in 2036 are propagated till the end of the simulation. 
%
\begin{table*}
	\centering
	\caption{Overall results for the ADP--IS method (order 8, tolerance $10^{-10}$, $N_{\text{max}}$ 12, $3\sigma$ domain).}
	\label{tab:ADP_3sigma}
	\begin{tabular}{ccccccc} 
		\hline
		$n_{\text{PHS}}$ & $t_{\text{ADP}}$ & $n_{\text{samples}}$ & $t_{\text{IS}}$ & $t_{\text{all}}$ & $\hat{p}$ & $\sigma(\hat{p})$ \\
		\hline
		267 & 4~h~6~min & 204293 & 26~min & 4~h~32~min & $1.17\cdot 10^{-5}$ & $2.93\cdot 10^{-6}$\\
		\hline
	\end{tabular}
\end{table*}
The pattern of subdomains is not altered by the introduction of the pruning. Simply, a large portion of the initial set is no longer investigated. This action has a strong impact on the number of propagated subdomains, that is now significantly lower (267). Consequently, the computational time required by the propagation reduces significantly (4~h~6~min). 

The pattern of generated subdomains represents the starting point for the second phase, the application of the IS method for the computation of the impact probability in 2036. We initialize the method by defining a uniform pdf including all the generated PHS's as ISD. The boundaries of the ISD on the $a-l$ plane are represented in blue in Fig.~\ref{fig:ADP_result}. Then, samples are drawn from the ISD and each sample is associated with a PHS if possible. For samples belonging to the PHS's, the state vector corresponding to the drawn sample at the truncation epoch of the related PHS is reconstructed, and a pointwise propagation up to the epoch of minimum geocentric distance is performed. Figure~\ref{fig:ADP_result_focus} shows a focus of the resulting subsets, whereas Fig.~\ref{fig:ADP-IS_result} shows the pattern of generated samples projected onto the $a-l$ plane. Samples belonging to the PHS's are represented in blue, whereas impacting samples are represented in yellow. Black dots represent discarded samples. Not all samples belong to the PHS's, due to the shape of the selected ISD. A uniform ISD over a domain of regular shape enclosing all PHS's represents the easiest choice and can be applied regardless the complexity of the PHS's pattern. On the other side, this selection leads to the black dots shown in Fig.~\ref{fig:ADP-IS_result}. These samples, however, have a minimal impact on the computational effort required by the method, as they are discarded as soon as they are identified. 

The selection of the IS method as sampling technique allows us to increase significantly the number of samples lying within the PHS's with respect to a standard Monte Carlo approach, and this advantage is made possible by the pruning action of the ADP propagation. The analysis of the distribution of the impacting samples on the $a-l$ plane, however, shows that these are confined to a limited region inside the PHS's. That is, not all PHS's actually give a contribution to the impact probability in 2036. This result is related to the selection of the amplitude of $\Delta T_{\text{ref}}$: the value was set in order to grant a conservative pruning action on the subsets. A more detailed analysis is offered in Section~\ref{Sub_DT_ref}.

\begin{figure}
	\includegraphics[width=\columnwidth]{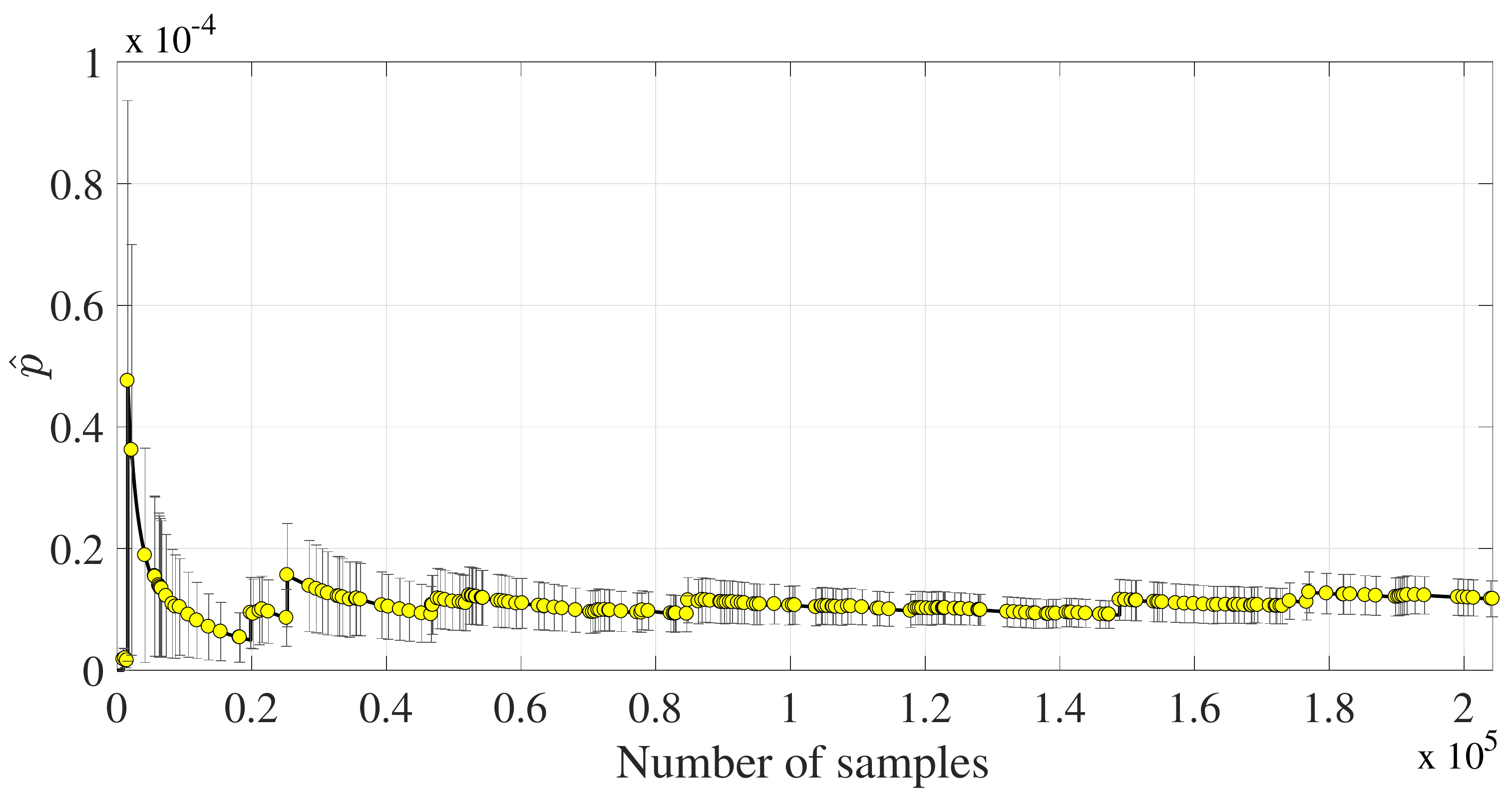}
    \caption{Estimated impact probability for Apophis resonant return 2036 as a function of the number of samples. In yellow, impacting samples. In grey, estimated Poisson statistics uncertainty ($1\sigma$).}
    \label{fig:P_est_trend}
\end{figure}

The trend of the estimated impact probability with the number of drawn samples is represented in Fig.~\ref{fig:P_est_trend}. Impacting samples are represented with yellow circles. The tolerance for the stopping criterion was set equal to $0.01$ per cent. After some initial significant oscillations, the impact probability asymptotically converges to the value of $1.17\cdot 10^{-5}$. The estimate is of the same order of magnitude of the reference value ($2.2\cdot 10^{-5}$) obtained with a standard Monte Carlo analysis, though slightly lower (see Section~\ref{Section_MC}). This difference can be explained considering the size of the propagated uncertainty set, as later explained in Section~\ref{Section_Uncertainty_size}.

An overview of the main results of the simulation is shown in Table~\ref{tab:ADP_3sigma}. Results are expressed in terms of number of PHS's $n_{\text{PHS}}$, computational time required by the ADP propagation $t_{\text{ADP}}$, number of generated samples for the IS method at convergence $n_{\text{samples}}$, computational time required by the IS method $t_{\text{IS}}$, overall computational time $t_{\text{all}}$, estimated impact probability value $\hat{p}$ and related Poisson statistics uncertainty $\sigma(\hat{p})$.

\section{Sensitivity analysis}
\label{Sensitivity}
The analysis presented in Section~\ref{Section_Apophis} was carried out starting from predefined values of expansion order, tolerance for the splitting routine, maximum number of splits and size of the uncertainty set. In this section, we investigate the role of the different parameters and provide some guidelines for the selection of the most appropriate set of parameters. The discussion is carried out dividing parameters mostly affecting the ADP propagation (order, tolerance, minimum box size and reference orbital period range) and parameters affecting the estimated impact probability (uncertainty box size).

\subsection{Selection of splitting tolerance, expansion order and $N_{\text{max}}$}
\label{Sub_eps}
As described in Section~\ref{Section_ADS}, the main parameters for the ADS propagation are the tolerance for the splitting procedure, the expansion order and the maximum number of splits. The selection of the tolerance is strictly related to the accuracy required in the description of the subsets at the end of the simulation. This concept is valid in both ADS and ADP propagation. Due to error accumulation during the integration process, indeed, the actual accuracy of the ADP result tends to decrease with respect to the imposed accuracy. This effect becomes more significant as the nonlinearities of the dynamics increase, so that, in order to grant a specific accuracy, the imposed tolerance must be in some cases some orders of magnitude lower. 

Table~\ref{tab:ADP_tolerance} shows the average accuracy in position for the subsets at the epoch of the first resonant return considering an expansion order equal to 8 and decreasing values of tolerance. We estimated the accuracy by comparing the results of pointwise propagations and polynomial evaluations for random samples drawn in the generated subsets. The error in position shown in Table~\ref{tab:ADP_tolerance} represents an average of the computed errors. As expected, there is a difference of around three orders of magnitude with respect to the imposed tolerance.

For the case under study, the error is strictly related to the intervening close encounter in 2029. As an example, if we perform an ADS propagation with order 8 and tolerance $10^{-10}$, and we stop the propagation three months before the close encounter in 2029, we obtain a position error of about $10^{-11}$ AU. This error expands to $10^{-8}$ AU six months later, i.e. three months after the close encounter. That is, the close encounter yields an increase of about 3 orders of magnitude in the position error. As the propagation continues, the error accumulates and reaches $5.95\cdot 10^{-7}$ AU at the epoch of the first resonant return. Therefore, the splitting tolerance is a critical parameter and its selection must account for all the above aspects. For our analysis, we selected a tolerance capable of granting a maximum error in position of 100 km. This requirement results into a splitting tolerance of at least $10^{-10}$.

\begin{table}
	\centering
	\caption{Average error in position as a function of the imposed tolerance for subsets at the epoch of the first resonant return (ADP propagation, order 8, $N_{\text{max}}$ 12, $3\sigma$ initial uncertainty set).}
	\label{tab:ADP_tolerance}
	\begin{tabular}{cc} 
		\hline
		Tolerance & Error \\
		\hline
		$10^{-8}$ & $4.83\cdot 10^{-6}$AU\\
		$10^{-9}$ & $1.24\cdot 10^{-6}$AU\\
		$10^{-10}$ & $5.95\cdot 10^{-7}$AU\\
		\hline
	\end{tabular}
\end{table}

\begin{table*}
	\centering
	\caption{Performance of the ADP--IS method for different expansion orders (tolerance $10^{-10}$, $N_{\text{max}}$ 12, $3\sigma$ initial uncertainty set).}
	\label{tab:ADP_order}
	\begin{tabular}{cccccccc} 
		\hline
		Order & $n_{\text{PHS}}$ & $t_{\text{ADP}}$ & $n_{\text{samples}}$ & $t_{\text{IS}}$ & $t_{\text{all}}$ & $\hat{p}$ & $\sigma(\hat{p})$\\
		\hline
		8 & 267 & 4~h~6~min & 204293 & 26~min & 4~h~32~min & $1.17\cdot 10^{-5}$ & $2.93\cdot 10^{-6}$\\
		
		7 & 267 & 1~h~53~min & 204293 & 24~min & 2~h~17~min & $1.17\cdot 10^{-5}$ & $2.93\cdot 10^{-6}$\\
		 
		6 & 267 & 49~min & 204293 & 23~min & 1~h~12~min & $1.17\cdot 10^{-5}$ & $2.93\cdot 10^{-6}$\\			
		5 & 267 & 24~min & 204293 & 23~min & 47~min & $1.17\cdot 10^{-5}$ & $2.93\cdot 10^{-6}$\\
		
		4 & 267 & 28~min & 204293 & 25~min & 53~min & $1.17\cdot 10^{-5}$ & $2.93\cdot 10^{-6}$\\
		3 & 589 & 8~min & 273035 & 9 h~47~min & 9~h~55~min & $8.60\cdot 10^{-6}$ & $2.20\cdot 10^{-6}$\\
		\hline
	\end{tabular}
\end{table*}

\begin{table*}
	\centering
	\caption{Performance of the ADP--IS method for different values of $N_{\text{max}}$ (order 5, tolerance $10^{-10}$, $3\sigma$ initial uncertainty set).}
	\label{tab:ADP_Nmax}
	\begin{tabular}{cccccccc} 
		\hline
		$N_{\text{max}}$ & $n_{\text{PHS}}$ & $t_{\text{ADP}}$ & $n_{\text{samples}}$ & $t_{\text{IS}}$ & $t_{\text{all}}$ & $\hat{p}$ & $\sigma(\hat{p})$\\
		\hline
		12 & 267 & 24~min & 204293 & 23~min & 47~min & $1.17\cdot 10^{-5}$ & $2.93\cdot 10^{-6}$\\
		
		11 & 148 & 17~min & 204293 & 26~min & 43~min & $1.17\cdot 10^{-5}$ & $2.93\cdot 10^{-6}$\\
		
		10 & 84 & 12~min & 204293 & 29~min & 41~min & $1.17\cdot 10^{-5}$ & $2.93\cdot 10^{-6}$\\			
		9 & 47 & 9~min & 204293 & 34~min & 43~min & $1.17\cdot 10^{-5}$ & $2.93\cdot 10^{-6}$\\
		
		\hline
	\end{tabular}
\end{table*}

\begin{table}
	\centering
	\caption{Trend of the computational times $t_{\text{ADP}}$ and $t_{\text{IS}}$ for increasing ($\uparrow$) values of the expansion order and minimum box size.}
	\label{tab:ADP-IS_trends}
	\begin{tabular}{ccc} 
		\hline
		& ADP & IS \\
		\hline
		Order $\uparrow$ & $t_{\text{ADP}} \uparrow \downarrow$ & $t_{\text{IS}} \uparrow \downarrow$\\
		$N_{\text{max}}$ $\uparrow$ & $t_{\text{ADP}} \uparrow $ & $t_{\text{IS}} \downarrow$\\
		\hline
	\end{tabular}
\end{table}

Expansion order and minimum box size, instead, play quite different roles in ADS and ADP propagation. During a DA propagation, a reduction of the expansion order causes a decrease in the accuracy of the results at a specific integration epoch. This decreased accuracy yields an increase in the required number of splits during the ADS propagation and, overall, a larger number of generated subsets. The role of the minimum box size, instead, is to limit the number of splits, so that, overall, both parameters have a strong influence on the number of generated subdomains and, as a consequence, on the required computational effort. The role of the expansion order is twofold, since a decrease in the order causes the number of subdomains to increase, but reduces the computational time required to perform a single integration step. Thus, it is reasonable to imagine that there exists a specific expansion order capable of minimizing the computational effort required by the ADS propagation. This value, obviously, changes according to the specific case under study. The role of the minimum box size, instead, is univocal: by increasing the value of $N_{\text{max}}$, the computational effort required by the ADS propagation increases.

In the case of ADP propagation, the analysis is quite different. The role of the two parameters for the two phases is reported in Table~\ref{tab:ADP-IS_trends}. More specifically, the ADP propagation aims to select only subsets whose integration stops before the resonant return of interest having reached their minimum box size. A change in the expansion order modifies the splitting history, which could, but not necessarily would, modify the overall number of splits. This behaviour has a direct impact on the required computational time, though the description of the role of the expansion order is not immediate. A decrease in the expansion order, indeed, may cause just earlier splits performed with the same splitting sequence, or a complete change in the splitting history. In the first case, the role of the expansion order becomes univocal: a reduction in the expansion order causes a decrease in the computational effort. In the second case, the changes in the splitting history and the number of generated subsets may be so relevant that what is gained in performing single integration steps may be lost in the longer propagation of the generated subsets. Overall, the role of the order in not univocal, and it exists an order that minimizes the computational time required by the ADP propagation. 

The role of the minimum box size, instead, is the same as in the ADS propagation: a decrease in the value of $N_{\text{max}}$ causes an earlier stop of the propagation of the subsets, and a reduction of the computational effort.

The whole procedure, however, includes both an ADP propagation and a sampling phase, and the role played by the two parameters during the sampling phase is different from the ADP phase. The role of the expansion order is, again, twofold: a reduction of the order causes longer pointwise propagations, but faster polynomial evaluations. The relative weight of the two effects essentially depends on the number of required samples. The role of the minimum box size, instead, is univocal and opposite with respect to the ADP propagation: a reduction of the value of $N_{\text{max}}$ implies longer pointwise propagations.

The selection of the best combination of order and minimum box size, therefore, relies on all these aspects. Starting from these considerations, we performed a sensitivity analysis in order to quantify the impact of the two parameters on the performance of the ADP--IS method for the case of the first resonant return of asteroid Apophis.

The results of a sensitivity analysis on the expansion order are shown in Table~\ref{tab:ADP_order}, considering six different expansion orders. The comparison is performed by considering the same parameters of the analysis presented in Section~\ref{Section_Apophis}. The second column shows the number of generated subdomains. The value is not affected by the expansion order till order 4, while for order 3 the value is more than doubled. This trend can be explained looking at the splitting history. For orders from 8 to 4, no split occurs before the close encounter in 2029, and the sequence of splits is exactly the same, though single splits are performed at different epochs. Things completely change with order 3, with 4 splits occurring before the 2029 close encounter. This change has a direct impact on the number of generated subsets. A difference can be detected also by looking at the required number of samples or at the estimated impact probability and related Poisson statistics uncertainty. Assuming not to alter the sequence of generated samples, values obtained with orders 8 to 4 are identical, whereas values obtained with order 3 are slightly different.

\begin{figure}
	\includegraphics[trim=0cm 0cm 0cm 0cm, clip=true, width=\columnwidth]{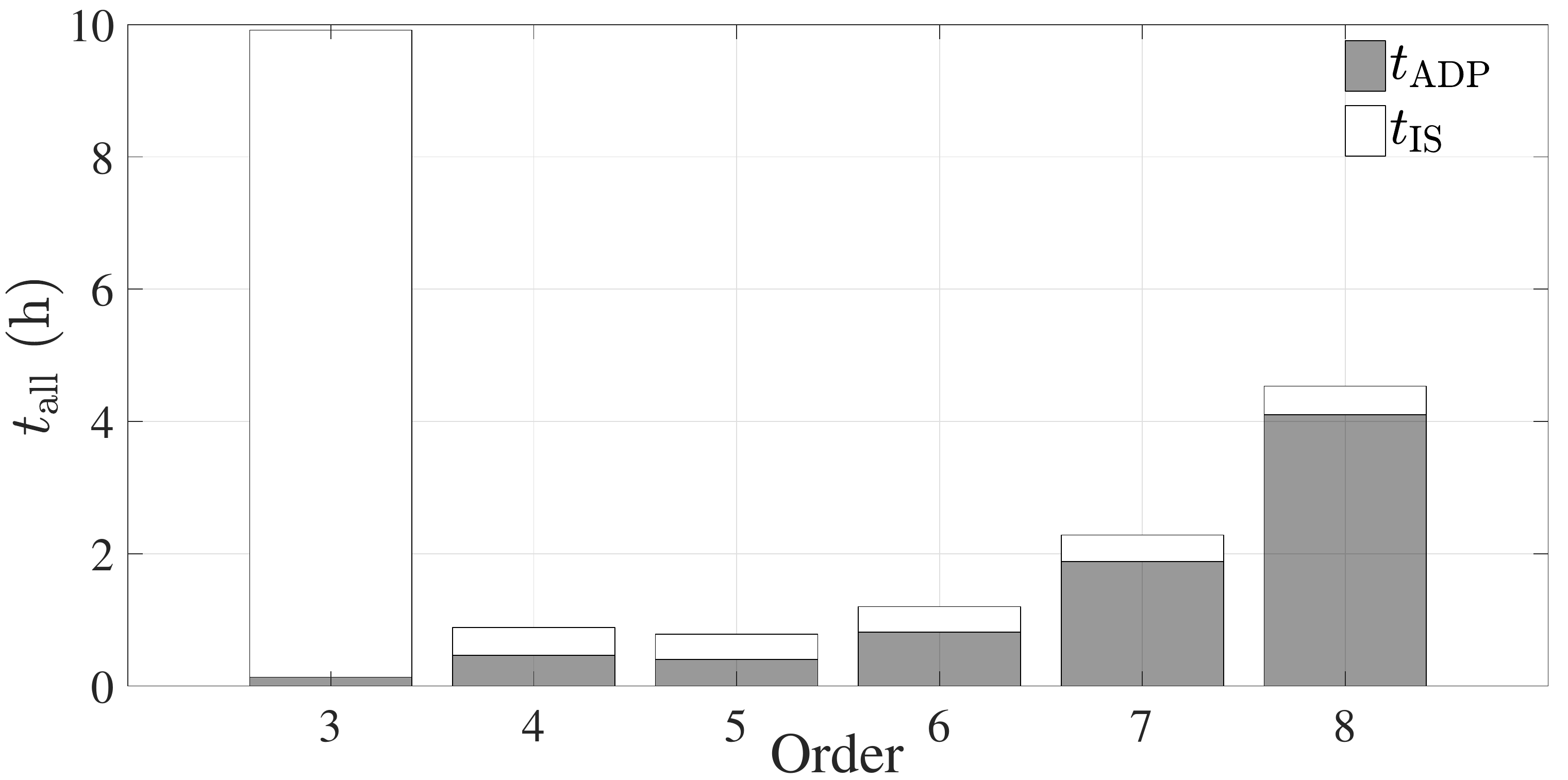}
    \caption{Computational time versus expansion order (ADP--IS method, tolerance $10^{-10}$, $N_{\text{max}}$ 12, $3\sigma$ uncertainty domain). Grey bars and white bars represent computational times required by the ADP propagation and IS phase respectively.}
    \label{fig:ADP_tCPU}
\end{figure}

The expansion order has a significant impact on the computational effort required by the ADP and sampling phases (columns 3 and 5). As expected, the trends are not monotonic. Thus, it is possible to identify order 5 as the expansion order capable of minimizing the overall computational time. Once again, it is interesting to see what happens with order 3: the early splits at the epoch of the 2029 close encounter completely change the splitting history, causing subsets to stop much earlier than what happens with larger orders. Unlike orders from 8 to 4, where the subsets are stopped few days before the expected resonant return in 2036, with order 3 the subsets are stopped around 2030, 6 years earlier. This earlier stop grants computational time saving for the ADP propagation, but has a tremendous backlash for the sampling phase, with each sample propagated for years instead of days. For this reason, the computational time required for the sampling phase is much larger (column 5). The trend of the computational time for the different orders is shown in Fig.~\ref{fig:ADP_tCPU}.

We limited the analysis in Table~\ref{tab:ADP_order} to order 3 as the minimum order, as early splits that appear with this order magnify with order 2, leading to 53 subsets generated before the 2029 close encounter. This behaviour exacerbates the limitations previously pointed out for low orders. Moreover, the error estimation procedure described in Section~\ref{Section_ADS} does not work with linear approximation and tends to provide inaccurate estimates with order 2.

We performed a similar analysis by considering the effect of the minimum box size on the required computational effort and estimated impact probability. Table~\ref{tab:ADP_Nmax} shows the results of the analysis considering the optimal expansion order identified in Table~\ref{tab:ADP_order} and the same values of tolerance and uncertainty box size of the previous simulations. As described before, the role of the minimum box size is univocal in the two phases, though opposite, and this trend is confirmed by the analysis: a decrease in the value of $N_{\text{max}}$ causes a reduced computational effort required by the ADP propagation but longer pointwise propagations for all samples. For the case under study, $N_{\text{max}}$ equal to 10 allows us to minimize the required computational effort.  

As in the previous case, a change in the value of $N_{\text{max}}$ does not alter the estimated impact probability, though the pattern of generated subsets is now modified. This trend can be explained considering the fact that, a reduction of the value of $N_{\text{max}}$ generates larger subsets at earlier truncation epochs, but with the same accuracy. Thus, if the drawn samples are fixed, their mapping to the epoch of the first resonant return is essentially the same. That is, the pattern of impacting samples is not altered.

As described in the presented analysis, both expansion order and minimum box size influence the performance of the method. In particular, the expansion order plays a key role in the definition of the computational effort required by the method, while the minimum box size has a lower influence. As the method is composed by two phases, we can say that the selection of the order must be done in order to minimize the computational effort required by the heaviest one. In our method, the ADP propagation plays this role, so that, in order to decrease its impact on the overall computational time, the most effective way is to reduce the expansion order, still limiting as much as possible the number of generated subsets before the first close encounter.

\subsection{Definition of $\Delta T_{\text{ref}}$}
\label{Sub_DT_ref}

\begin{table*}
	\centering
	\caption{Performance of the ADP--IS method for different values of $\varepsilon$ (order 5, tolerance $10^{-10}$, $N_{\text{max}}$ 10, $3\sigma$ initial uncertainty set).}
	\label{tab:ADP_eps}
	\begin{tabular}{cccccccccc} 
		\hline
		$\varepsilon$ & $n_{\text{D}}$ & $n_{\text{PHS}}$ & $n_{\text{imp}}$ & $t_{\text{ADP}}$ & $n_{\text{samples}}$ & $t_{\text{IS}}$ & $t_{\text{all}}$ & $\hat{p}$ & $\sigma(\hat{p})$\\
		\hline
		0 & 71 & 71 & 43 & 11~min & 204293 & 29~min &  40~min & $1.17\cdot 10^{-5}$ & $2.93\cdot 10^{-6}$\\
		
		$10^{-5}$ & 71 & 71 & 43 & 11~min & 204293 & 29~min &  40~min & $1.17\cdot 10^{-5}$ & $2.93\cdot 10^{-6}$\\
		
		$10^{-3}$ & 84 & 84 & 43 & 12~min & 204293 & 29~min & 41~min & $1.17\cdot 10^{-5}$ & $2.93\cdot 10^{-6}$\\			
		$1$ (ADS) & 470 & 121 &  41 & 1~h~31~min & 216299 & 31~min & 2~h~02~min & $1.11\cdot 10^{-5}$ & $2.25\cdot 10^{-6}$\\
		
		\hline
	\end{tabular}
\end{table*}

As described in Section~\ref{Section_ADP}, the reference orbital period range $\Delta T_{\text{ref}}$ represents the key parameter for the ADP propagation. The range, centred in the selected resonant return period $T'$, is defined to account both inaccuracies in the estimation of the orbital period range of the subdomains and small dynamical perturbations between the first close encounter and the predicted resonant return. The semi-amplitude of this range, $\varepsilon$, plays therefore a key role as it influences both the accuracy of the probability estimates and the required computational time.

Table~\ref{tab:ADP_eps} shows the performance of the ADP--IS method for different values of the reference range semi-amplitude $\varepsilon$. With respect to the previous analyses, two additional parameters are shown: the number of generated subdomains $n_D$ and the number of subdomains that include impacting samples $n_{\text{imp}}$. These two parameters provide, along with the number of PHS's $n_{\text{PHS}}$, a clear picture of the pruning action performed during the ADP propagation. We performed the analysis considering four values of $\varepsilon$. In particular, it is interesting to analyse what happens considering the two limiting cases: $\varepsilon=0$ and $\varepsilon=1$. 

In the first case, the reference orbital period range collapses to the value of $T'$, i.e. subsets are maintained throughout the simulation only if their estimated orbital period range includes $T'$. In this case, the number of PHS's is lower than the one obtained with $\varepsilon=10^{-3}$, but the value of impact probability is exactly the same. This result can be explained considering that the pattern of impacting samples is not altered, as confirmed by the parameter $n_{\text{imp}}$. That is, for the case under study, a less conservative selection of the parameter $\varepsilon$ would allow us to obtain the same results, though no evident savings in computational time would be obtained. The selection of $\varepsilon=0$ corresponds to considering a Keplerian motion between the two encounters. As described in \citet{Valsecchi2003}, this assumption may provide quite accurate results for the timing, and for the case under study, where no significant perturbations between the two encounters exist, it can be considered acceptable.

Let us now analyse the second limit case. If $\varepsilon$ is set to $1$, we are essentially selecting a very large reference range for the orbital period. That is, the ADP propagation becomes an ADS propagation, i.e. no pruning is performed and all subsets are propagated until the final simulation time or the maximum number of splits is reached. The sampling phase, instead, is not significantly altered: the ISD is defined including subsets whose propagation stops before the expected resonant return. Results are reported in the last row of Table~\ref{tab:ADP_eps}. The value of impact probability is similar to the one obtained with the pruning action, but the number of generated subsets is much larger, which affects in turn the required computational time. That is, a very conservative selection of $\varepsilon$ would yield a factor three increase in computational time.

The selection of the parameter $\varepsilon$ is therefore crucial. Within the assumptions of our method, i.e. small perturbations between the two encounters, we can define an upper threshold for the $\varepsilon$ value as
\begin{equation}
    h\varepsilon T'v_{\earth}<0.05\,\text{AU}
	\label{eq:IVP}
\end{equation}
where $h$ is the number of revolutions of the asteroid between the encounters, whereas $v_{\earth}$ is the Earth heliocentric velocity. The expression on the left hand side of the inequality represents the heliocentric arc covered by the Earth in the time range $h\varepsilon T'$. When we define the reference range $\Delta T_{\text{ref}}$ and we compare it with $\Delta T_{\text{sub}}$ of a given subset, therefore, we are verifying that the uncertainty in the position of the current subset with respect to the Earth position is lower than 0.05 AU, i.e. the current subset can be labelled as Potentially Hazardous. For the case under study, this value is about $10^{-3}$, the value selected for the analysis presented in the previous sections. 

As previously stated in the paper, the application of the ADP--IS method is strictly related to the assumption of no intervening close approaches with other major bodies in the investigated time window. In case of expected close approaches, indeed, the situation drastically changes as the resonances estimated at the epoch of the first close encounter may lose their validity. In such cases, one must rely on the more conservative approach of ADS propagation for investigating a selected propagation window, obtaining accurate results with unavoidable drawbacks in efficiency.  

The decision whether to perform an ADP propagation or disable pruning is made based on a preliminary analysis of the possible trajectories of the asteroid between the two encounters. For the case under study, the availability of the dispersion of Apophis' orbital parameters after the first close encounter allows us to estimate the minimum orbit intersection distance (MOID) dispersion between the asteroid and the other main bodies of the Solar System (see \citet{Armellin2010a}). This fast survey provides us with an overview of possible close approaches between the two encounters and drives our decision on the propagation method. For the case under study, the analysis required less than one minute and allowed us to exclude any significant encounter with other planets in between.

\subsection{Effect of the size of the uncertainty domain}
\label{Section_Uncertainty_size}
\begin{figure*}
\centering
\subfloat[\label{subfig:3s_vs_4s}]
{\includegraphics[width=\columnwidth]{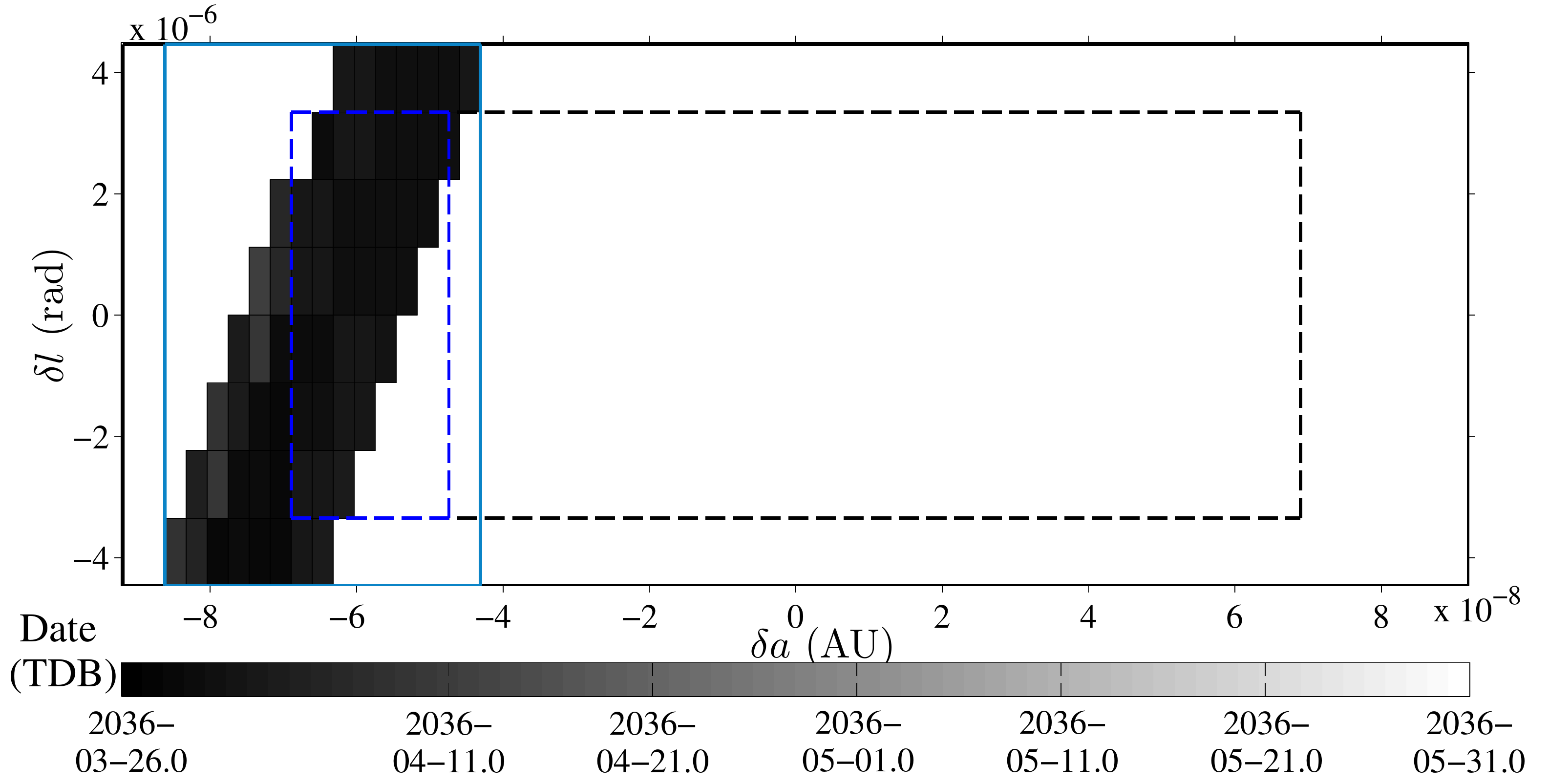}
}
\hfill
\subfloat[\label{subfig:3s_vs_5s}]
{\includegraphics[width=\columnwidth]{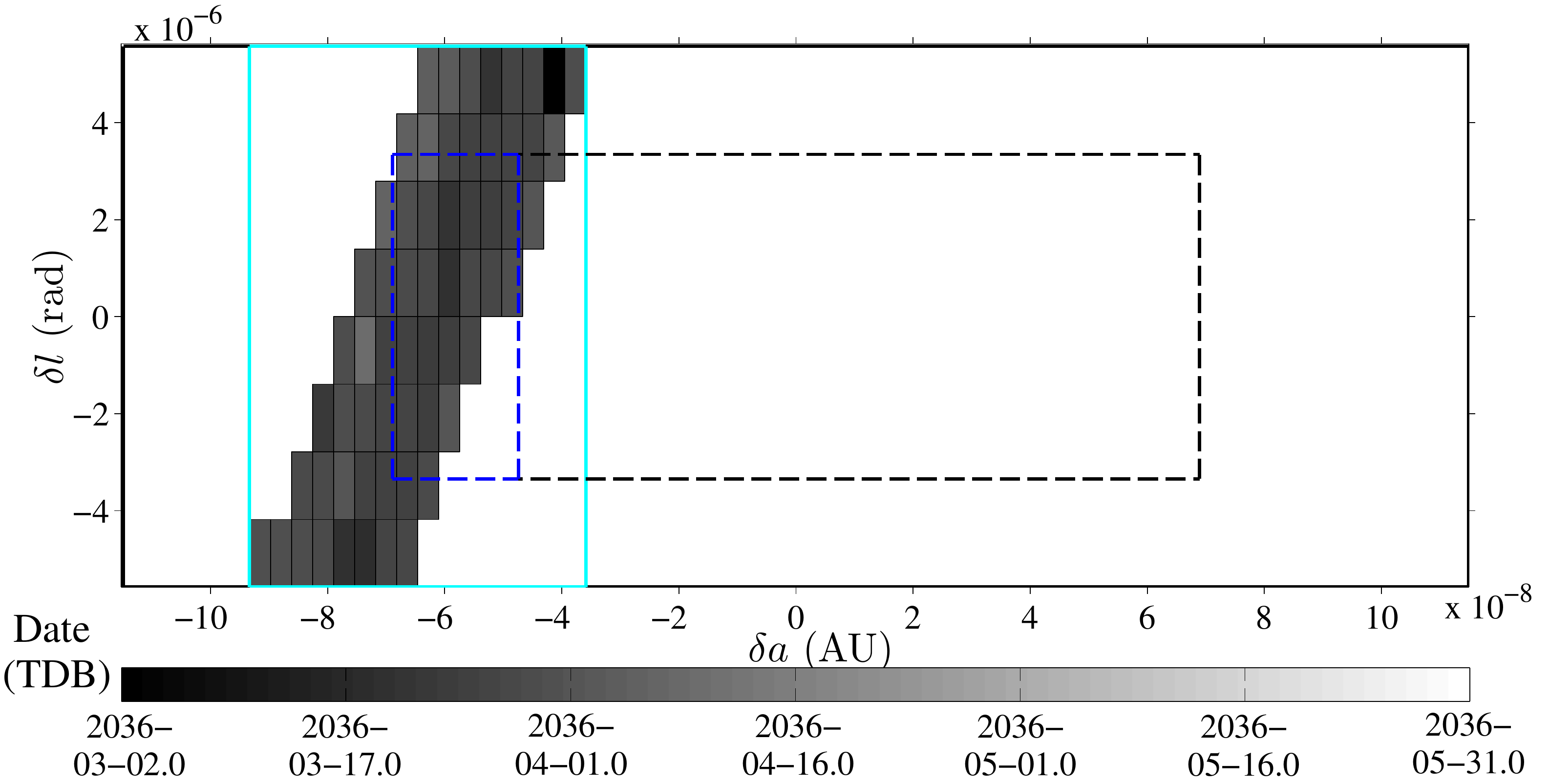}
}
\caption{Comparison between $4\sigma$ \protect\subref{subfig:3s_vs_4s} and $5\sigma$ \protect\subref{subfig:3s_vs_5s} cases. The figure box and the dashed black lines represent the selected boundaries ($4\sigma$ in \protect\subref{subfig:3s_vs_4s} and $5\sigma$ in \protect\subref{subfig:3s_vs_5s}) and the $3\sigma$ boundaries respectively. In light blue and cyan, the ISD boundaries for the $4\sigma$ and $5\sigma$ respectively. In dashed blue, the $3\sigma$ ISD boundaries.}
\label{fig:Samples}
\end{figure*}

\begin{figure*}
	\includegraphics[trim=0cm 0cm 0cm 0cm, clip=true,width=0.6\textwidth]{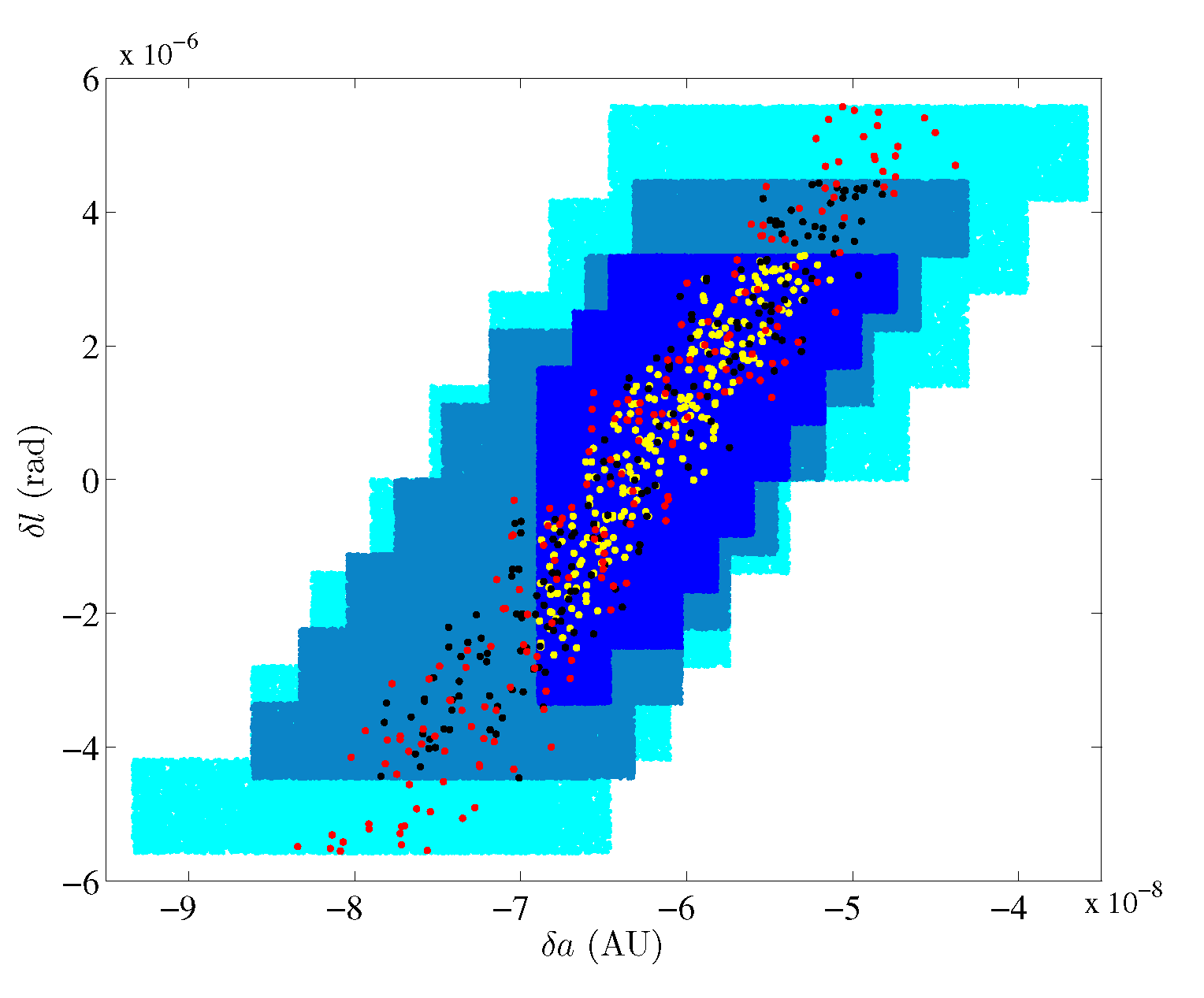}
    \caption{Comparison of the samples distribution for the $3\sigma$, $4\sigma$ and $5\sigma$ cases. In blue, light blue and cyan, drawn samples for the $3\sigma$, $4\sigma$ and $5\sigma$ case respectively. In yellow, black and red, impacting samples for the $3\sigma$, $4\sigma$ and $5\sigma$ case respectively.}
    \label{fig:All_samples}
\end{figure*}

\begin{figure*}
\centering
\subfloat[\label{subfig:Samples_3s_vs_4s}]
{\includegraphics[width=\columnwidth]{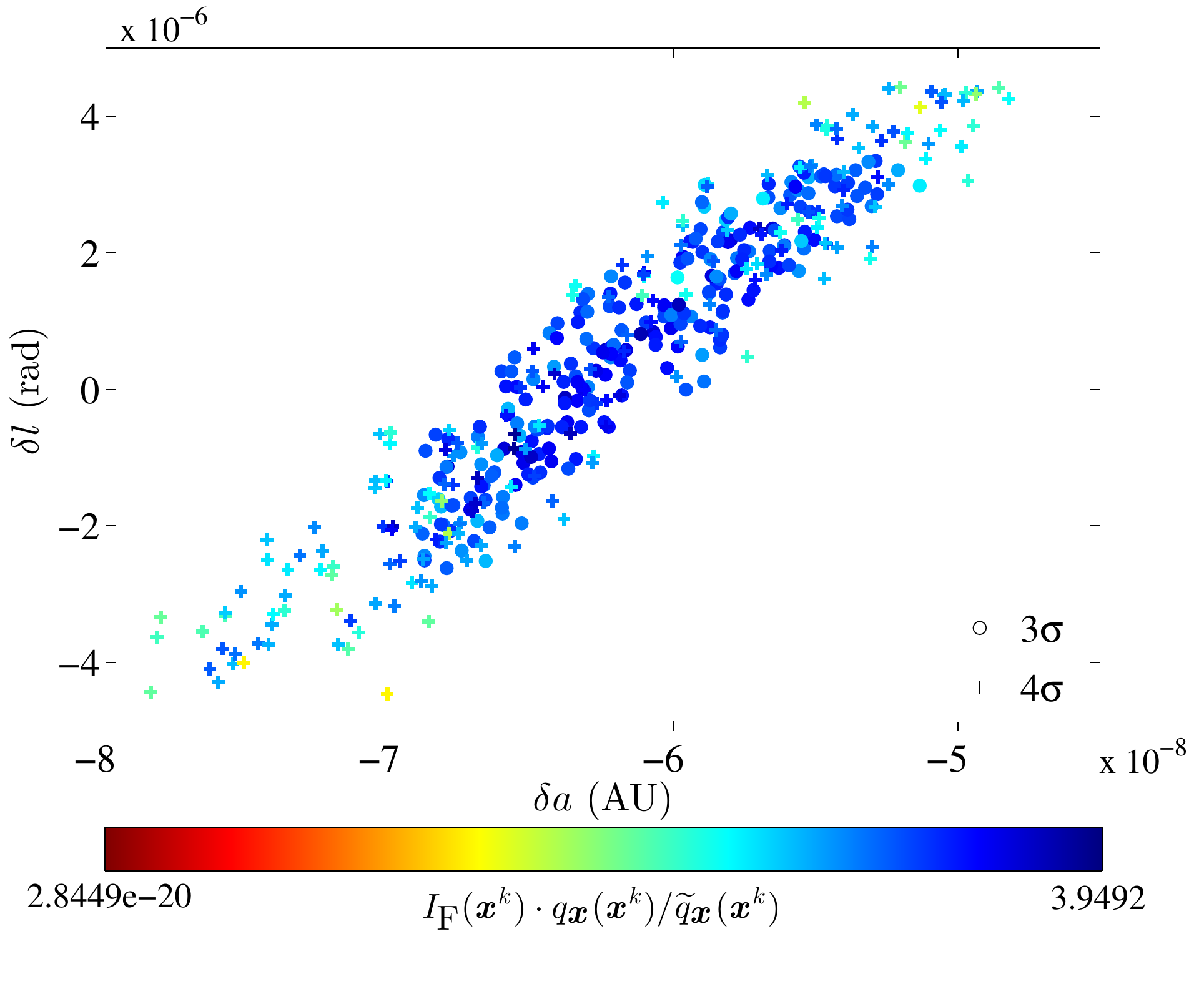}
}
\hfill
\subfloat[\label{subfig:Samples_3s_vs_5s}]
{\includegraphics[width=\columnwidth]{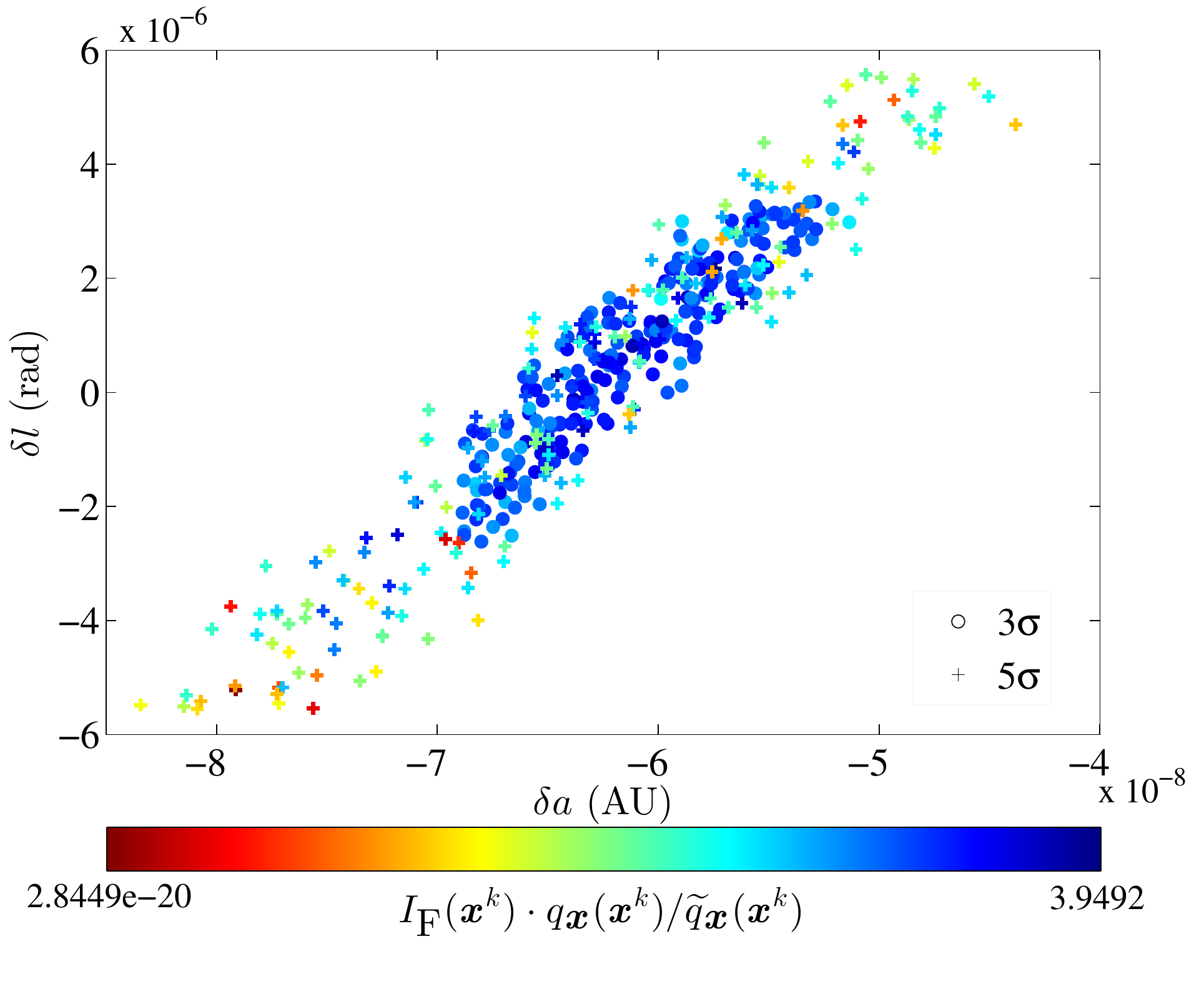}
}
\caption{Impacting samples distribution for $4\sigma$ \protect\subref{subfig:Samples_3s_vs_4s} and $5\sigma$ \protect\subref{subfig:Samples_3s_vs_5s} cases. Colours are referred to the associated contribution to the impact probability.}
\label{fig:Samples}
\end{figure*}

The analysis presented in the previous sections was done considering different values of expansion order, tolerance for the splitting procedure and minimum box size, whereas we considered only one size for the initial uncertainty set, that is a $3\sigma$ 6-dimensional rectangle. This selection, in a 6-dimensional problem with uncorrelated variables, consists in considering the 98.4 per cent of the probability mass. In the following paragraphs, we present the impact of the size of the uncertainty set on the results of the ADP--IS method. It is worth noting that, as previously mentioned in Section~\ref{Section_Apophis}, in case of full covariance matrix, the initial uncertainty box would be defined in the eigenvector space in order to avoid wrapping effect and including very low probability solutions, so all the analyses and values presented in this section hold for the more general case of correlated variables.

The ADP propagation and the IS phase are strongly influenced by the selection of the size of the initial uncertainty set. The ADP propagation, indeed, limits the generation of the subsets within the boundaries of the considered uncertainty set, and this aspect influences also the shape of the ISD for the impact probability computation phase. Samples, indeed, are confined within the initial uncertainty set, impacting samples are found only within these limits and possible impacting samples that lie out of the initial uncertainty set are discarded. This aspect distinguishes our sampling approach from a standard Monte Carlo method, where samples are drawn directly from the original probability density function, and the probability of drawing samples is determined by the pdf itself. In principle, samples could lie anywhere in the uncertainty region.

Starting from these considerations, it is therefore interesting to study how the estimate for the impact probability changes with an increasing size of the initial uncertainty set. We initially performed the analysis by considering a size for the initial uncertainty set of $3\sigma$ (i.e. what was previously presented), $4\sigma$ and  $5\sigma$. 

Figure~\ref{subfig:3s_vs_4s} shows the results for the ADP propagation considering order 5, tolerance $10^{-10}$, $N_{\text{max}}$ equal to 10 and an initial uncertainty set size of $4\sigma$. The ISD boundaries on the $a-l$ plane are represented in light blue. On the same plot, the $3\sigma$ boundaries are represented with black dashed lines, whereas the ISD boundaries for the $3\sigma$ case are represented with dashed blue lines. The plot allows us to compare the sample regions in the two cases. In particular, a portion of the PHS's generated during the $4\sigma$ ADP propagation is not considered during the $3\sigma$ case. The $5\sigma$ case is shown in Fig.~\ref{subfig:3s_vs_5s}, representing the $5\sigma$ ISD in cyan.

Figure~\ref{fig:All_samples} shows a comparison of the results of the sampling phase for the $3\sigma$, $4\sigma$ and $5\sigma$ cases. Blue points represent drawn samples belonging to the $3\sigma$ case, with impacting samples represented as yellow dots. Light blue points represent drawn samples belonging to the $4\sigma$ case, with impacting samples represented as black dots. Finally, cyan points represent accepted samples for the $5\sigma$ case, with impacting samples represented with red dots. The analysis of the plot offers a clear picture of how the sampling region changes in the three cases. Moreover, it is possible to see how the increasing size of the initial uncertainty set allows us to include impacting samples out of the $3\sigma$ domain. While a $3\sigma$ domain appears as a too narrow selection, the $4\sigma$ and $5\sigma$ domains offer a better description of the impact region. Figures~\ref{subfig:Samples_3s_vs_4s} and~\ref{subfig:Samples_3s_vs_5s} show the distribution of the impacting samples for the three different simulations, with colors showing the contribution to the overall impact probability.

Table~\ref{tab:ADP_domain} shows the results of the analysis, including the $6\sigma$ case. With reference to the previous analyses, we added the parameter $p_{\text{out}}$, which represents the probability mass outside the selected uncertainty set, i.e. the complementary to 1 of the integral of the pdf over the considered domain. We remark that, because we use rectangular uncertainty sets, the values of $p_{\text{out}}$ shown in Table~\ref{tab:ADP_domain} are significantly smaller than those corresponding to the more commonly used ellipsoidal uncertainty regions, for which $p_{\text{out}}$ is equal to $0.17$, $1.38\cdot10^{-2}$, $3.41\cdot10^{-4}$ and $2.76\cdot10^{-6}$ for the $3\sigma$, $4\sigma$, $5\sigma$ and $6\sigma$ cases respectively.

By increasing the size of the initial uncertainty set, the computational time required by the ADP propagation increases. This trend is essentially due to the larger computational effort required by a single integration step and the larger number of subsets. An increase in the number of generated PHS's can be detected passing from $3\sigma$ to $4\sigma$, whereas this value remains essentially the same in the $5\sigma$ and $6\sigma$ cases.

\begin{table*}
	\centering
	\caption{Performance of the ADP--IS method for different values of initial uncertainty domain size (order 5, tolerance $10^{-10}$, $N_{\text{max}}$ equal to 10).}
	\label{tab:ADP_domain}
	\begin{tabular}{ccccccccc} 
		\hline
		Domain & $p_{\text{out}}$ & $n_{\text{PHS}}$ & $t_{\text{ADP}}$ & $n_{\text{samples}}$ & $t_{\text{IS}}$ & $t_{\text{all}}$ & $\hat{p}$ & $\sigma(\hat{p})$\\
		\hline
		$3\sigma$ & $1.61\cdot 10^{-2}$ & 84 & 12~min & 204293 & 29~min & 41~min & $1.17\cdot 10^{-5}$ & $2.93\cdot 10^{-6}$\\
		
		$4\sigma$ & $3.80\cdot 10^{-4}$ & 107 & 21~min & 300643 & 39~min & 1~h & $1.61\cdot 10^{-5}$ & $5.70\cdot 10^{-6}$\\
		
		$5\sigma$ & $3.44\cdot 10^{-6}$ & 104 & 29~min & 341804 & 41~min & 1~h~10~min & $1.90\cdot 10^{-5}$ & $6.81\cdot 10^{-6}$\\
						
		$6\sigma$ & $1.18\cdot 10^{-8}$ & 107 & 46~min & 353056 & 42~min & 1~h~28~min & $2.09\cdot 10^{-5}$ & $7.00\cdot 10^{-6}$\\
			
		\hline
	\end{tabular}
\end{table*}

The analysis of the sampling phase shows some interesting results. As expected, an increase in the initial uncertainty size causes an increase in the estimated impact probability value. Essentially, regions of the uncertainty set that were not studied during the ADP propagation for the $3\sigma$ case are now considered, and impacting samples can be found also in these regions. As a result, the estimated impact probability values for the $5\sigma$ and $6\sigma$ cases become very close to the reference value. The enlargement of the investigated region causes also an increase in the Poisson statistics uncertainty of the estimate. This result is expected too, as the variance is proportional to the sample region volume (see equation~(\ref{eq:IS_variance})). The analysis of the required number of samples at convergence shows that this value increases for larger initial uncertainty sets, and this trend reflects back on the computational time required by the sampling phase. 

The size of the uncertainty set should be selected to achieve the desired resolution on the impact probability, which is directly expressed by the parameter $p_{\text{out}}$. For the case under study, with an estimated impact probability of the order of $10^{-5}$, we selected a $6\sigma$ domain, which excludes only $1.18\cdot 10^{-8}$ of the probability mass.

\section{Comparison with standard and advanced orbital sampling techniques}
\label{Section_MC}

The analysis presented in the previous sections showed how the ADP--IS method represents a valuable tool for uncertainty propagation and impact probability computation for the first resonant return of a NEO. In order to assess the efficiency of the method with respect to other impact probability computation tools, we present in this section a comparison with standard and advanced orbital sampling techniques. In the first part, we compare our approach with Monte Carlo sampling techniques based on sample generation on the whole uncertainty set. Finally, we present a general comparison with the most used technique for impact probability calculation, the LOV method.

A first comparison can be done considering a standard Monte Carlo approach, where samples are drawn from the covariance matrix directly at the initial epoch (June 18, 2009). This method is probably the most straightforward approach but also the most expensive one, as the sampling is performed on the whole domain, and the propagation of each sample starts from the observation epoch. By performing the propagation of one million samples, the estimated impact probability results into $2.2\cdot 10^{-5}$ , whereas the Poisson statistics uncertainty is equal to $4.71\cdot 10^{-6}$. We selected the number of samples in order to detect a non null value of impact probability \citep{Farnocchia2015}.

Unfortunately, if the Monte Carlo simulation is performed considering the same conditions of our method (i.e same dynamics, single core), the required computational time is much larger. The average computational time required to perform a single pointwise propagation from the initial epoch to the epoch of the first resonant return is $\approx1.2$~s. As a result, within the computation time required by the ADP--IS method for the $6\sigma$ case (see Table~\ref{tab:ADP_domain}), about 4400 samples could be propagated, which is not enough to estimate the expected impact probability. All this would lead to an estimated computational time of around two weeks for propagating one million samples on a single core. This value is of course not realistic, as typically Monte Carlo analyses can be easily set up in a multi-thread environment, thus granting significant savings in computational time. It is interesting, however, to highlight the significant savings that our approach grants with respect to  standard MC approach in the same conditions. The ADP--IS method, indeed, employs a lower number of samples, as samples are drawn just in a subset of the uncertainty set. Moreover, the propagation of all samples starts immediately before the resonant return, while in a standard Monte Carlo approach each sample is propagated starting from June 18, 2009. Therefore, the computational effort required by the ADP propagation is largely repaid later by shorter pointwise propagations and a reduced number of samples.

\begin{figure*}
\centering
\subfloat[\label{subfig:SS_scheme1}]{\includegraphics[trim=3.5cm 2cm 5cm 3cm, clip=true, width=0.33\textwidth]{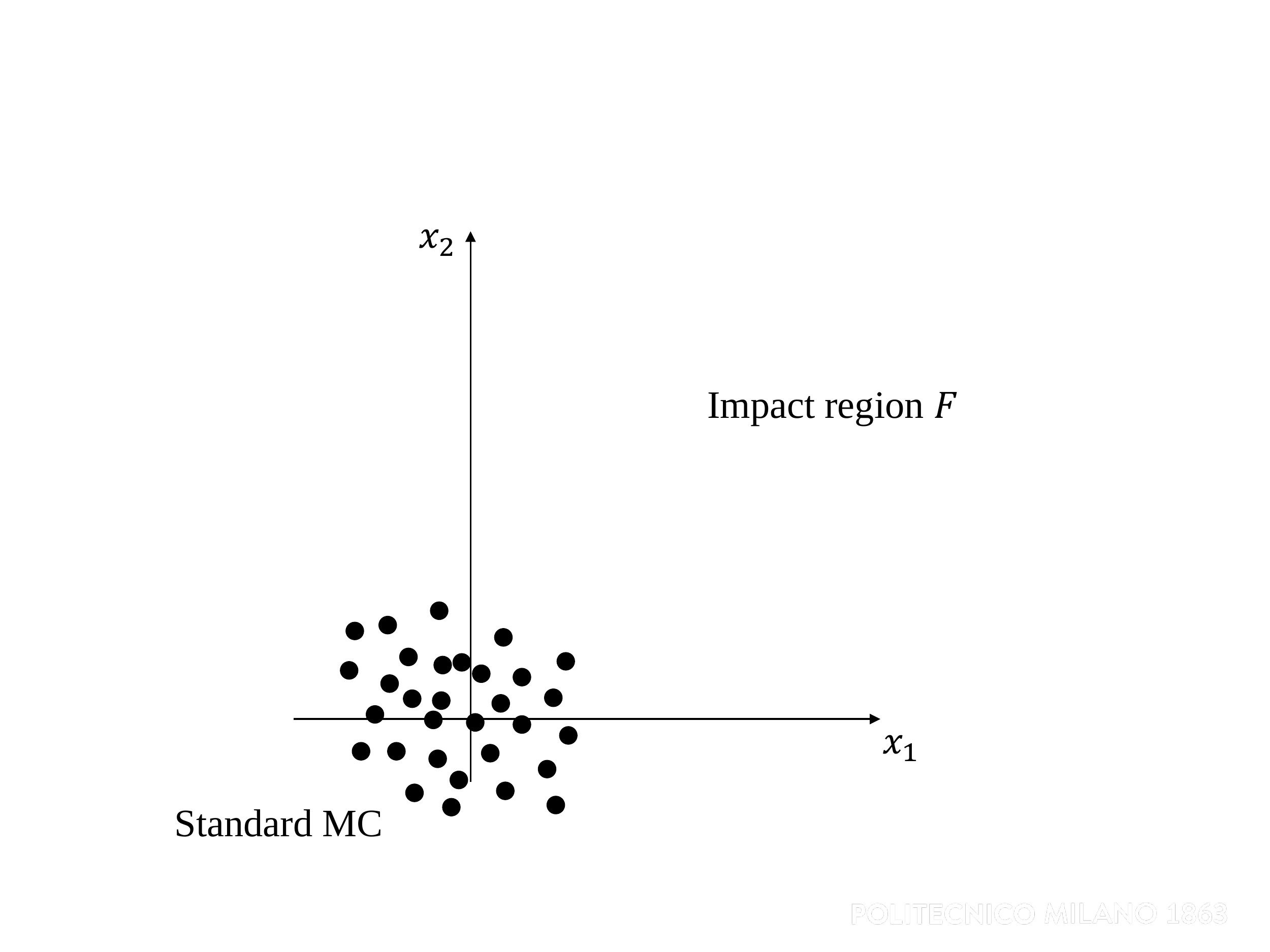}
}
\hspace{0cm}
\subfloat[\label{subfig:SS_scheme2}]{\includegraphics[trim=3.5cm 2cm 5cm 3cm, clip=true, width=0.33\textwidth]{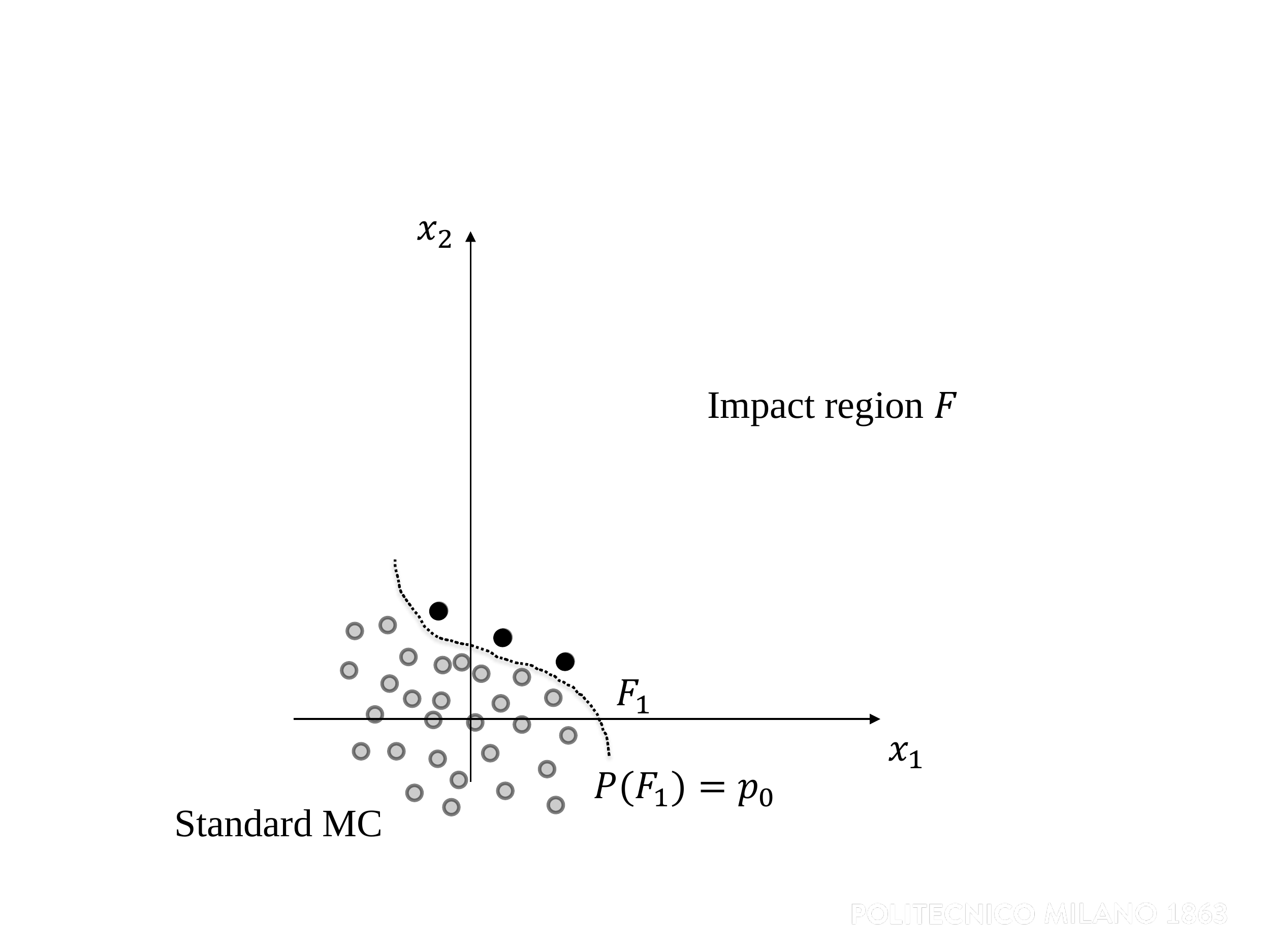}
}
\vspace{0cm}
\subfloat[\label{subfig:SS_scheme3}]{\includegraphics[trim=3.5cm 2cm 5cm 3cm, clip=true, width=0.33\textwidth]{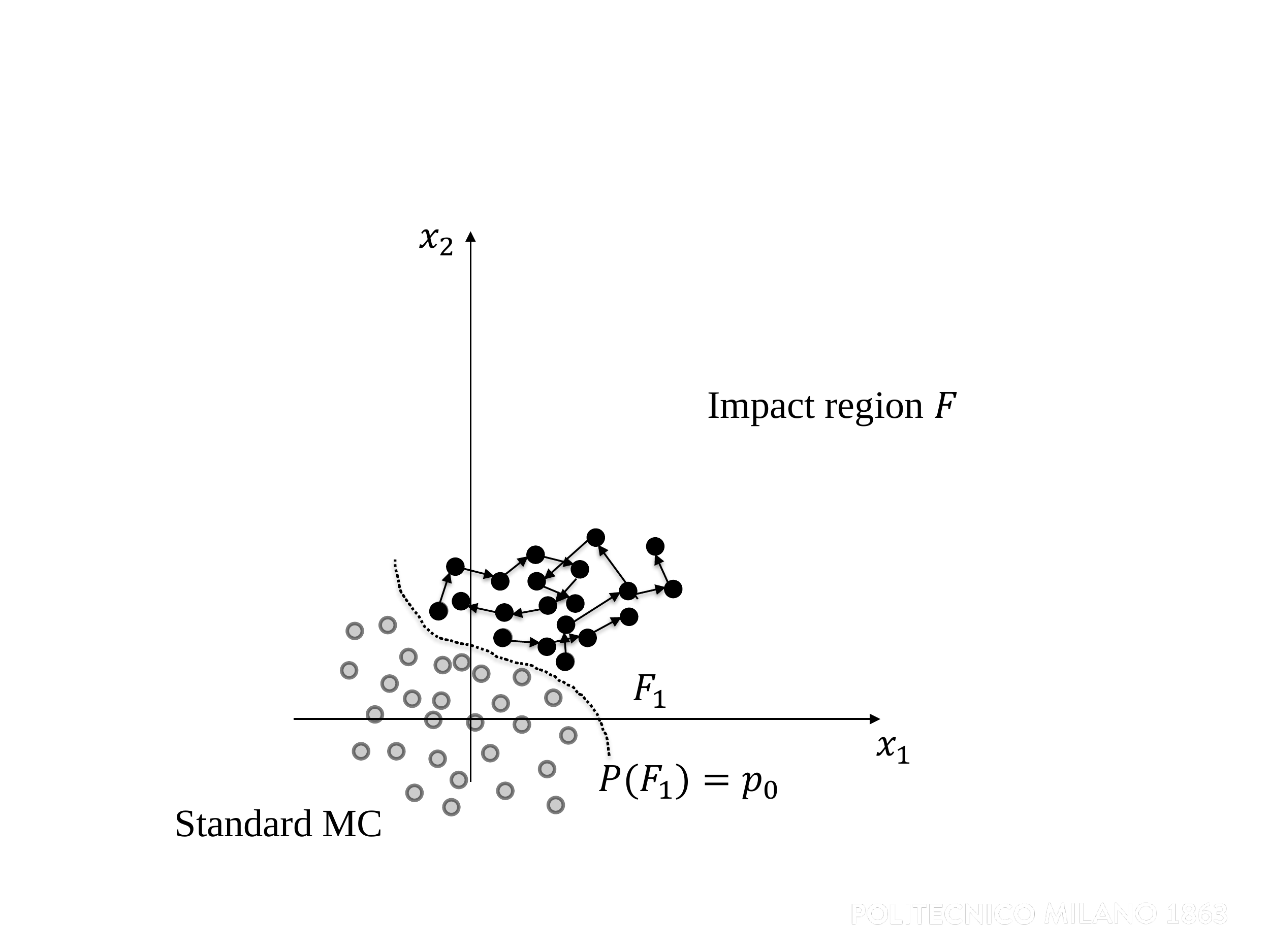}
}
\hspace{0cm}
\subfloat[\label{subfig:SS_scheme4}]{\includegraphics[trim=3.5cm 2cm 5cm 3cm, clip=true, width=0.33\textwidth]{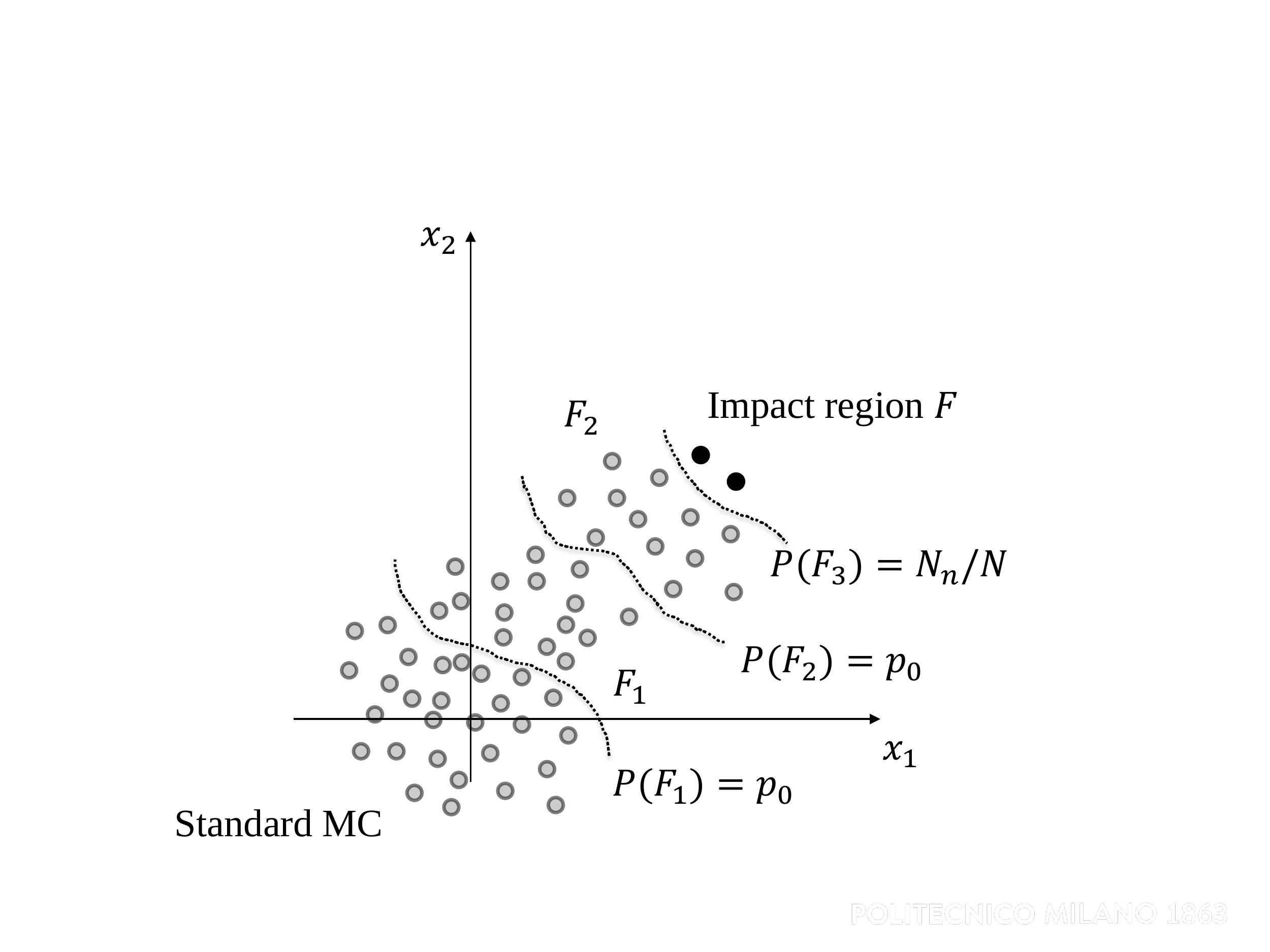}
}
\caption{Subset Simulation process: \protect\subref{subfig:SS_scheme1}, initialization by standard MC, \protect\subref{subfig:SS_scheme2}, CL 1 identification, \protect\subref{subfig:SS_scheme3}, samples generation by means of MCMC, \protect\subref{subfig:SS_scheme4}, new iterations and impact region identification.}
\label{fig:SS_scheme}
\end{figure*}

We show now a comparison with an advanced Monte Carlo technique called Subset Simulation (SS). The basic idea of SS is to compute small failure probabilities as the product of larger conditional probabilities \citep{Au2001,Zio2009,Zuev2012}. Given a target failure event $F$, let $F_1 \supset F_2 \supset ... \supset F_{n}=F$ be a sequence of intermediate failure events, so that $F_k=\bigcap\nolimits_{i=1}^{k}F_i,\quad k=1,2,...,n$. Considering a sequence of conditional probabilities, then the failure probability becomes:
\begin{equation}
	p(F)=p(F_1)\prod\limits_{i=1}^{n-1}p(F_{i+1}|F_{i})
	\label{eq:SS_pF}
\end{equation}
where $P(F_{i+1}|F_{i})$ represents the probability of $F_{i+1}$ conditional to $F_{i}$. A detailed description of the algorithm can be found in \citet{Au2001}. In the problem under study, the failure $F$ represents an impact with Earth, i.e. a geocentric distance smaller than the Earth radius. The method is initialized using standard MC to generate samples at the so-called conditional level (CL) 0 starting from the available nominal state vector and related uncertainty of the investigated object at the observation epoch. The number of samples generated at this level is maintained for each generated conditional level and it is referred to as $N$. Once the failure region $F_1$ is identified, a Monte Carlo Markov Chain (MCMC) Metropolis Hastings algorithm is used to generate conditional samples in the identified intermediate failure region. Another intermediate failure region is then located, and other samples are generated by means of MCMC. The procedure is repeated until the target failure region is identified. An illustration of the method is shown in Fig.~\ref{fig:SS_scheme}.

The approach was originally developed for the identification of structural failures, but it was also used in different research areas in reliability such as the definition of failure probabilities of thermo-hydraulic passive systems. The method was recently applied to the computation of space debris collisional probabilities by \citet{Morselli2014}.

In the presented approach, the intermediate failure regions are identified by assuming a fixed value of conditional probability $p_0=p(F_{i+1}|F_i)$. The identification of each conditional level, therefore, is strictly related to this value, and changes accordingly step by step, as explained in the followings. The resulting SS algorithm follows the general description presented in \citet{Morselli2014} and goes through the following steps:

\begin{enumerate}
\item Set $i=0$ and generate $N$ samples $\mathbfit{x}_0^{0,k},\quad k=1\,,\,...\,,\,N$ at  conditional level 0 by standard MC starting from the available state estimate of the investigated objects at the initial epoch $t_0$.

\item Propagate each sample up to the epoch of the first resonant return and compute its minimum geocentric distance. Note that, as in the ADP--IS method, the resonances can be easily determined by propagating the uncertainty set up to the epoch of the first close encounter by means of DA and evaluating the orbital period range.

\item Sort the $N$ samples in descending order according to the associated geocentric distance at the epoch of the first resonant return.

\item Identify an intermediate threshold value $D_{i+1}$ as the geocentric distance corresponding to the $(1-p_0)N$th element of the sample list. Define the $(i+1)$th conditional level as $F_{i+1}=\{d<D_{i+1}\}$, where $d$ represents the geocentric distance. According to the definition of $D_{i+1}$, the associated conditional probability $p(F_{i+1}|F_i)=p_0$.

\item If $D_{i+1}<R_{\earth}$, i.e. the geocentric threshold distance is lower than the Earth radius, go the the last step, otherwise select the last $p_0N$ samples of the list $\mathbfit{x}_0^{i,j}, \quad j=1,\ldots{},p_0N$. By definition, these samples belong to the $(i+1)$th conditional level.

\item Using MCMC, generate $(1-p_0)N$ additional conditional samples starting from the previously selected seeds belonging to $F_{i+1}$. A sample is set to belong to $F_{i+1}$ according to the following performance function:
\begin{equation}
    g_{\mathbfit{x}}^{i+1}(\mathbfit{x}_0)=d(\mathbfit{x}_0)-D_{i+1} \begin{cases}
    >0 \quad \mathbfit{x}_0\,\text{is\,out\,of\,the\,}(i+1)\,\text{th\,CL}\\
	=0 \quad \mathbfit{x}_0\,\text{is\,at\,the\,limit\,of\,the\,CL}\\
	<0 \quad \mathbfit{x}_0\,\text{belongs\,to\,the\,}(i+1)\,\text{th\,CL}
\end{cases}
\end{equation}
\item Set $i=i+1$ and return to step 2.

\item Stop the algorithm.

\end{enumerate}

The total number of generated samples is
\begin{equation}
	N_T=N+(n-1)(1-p_0)N
\end{equation}
where $n$ is the overall number of conditional levels required to reach the impact region. Since the conditional probability is equal to $p_0$ for each level, the impact probability expressed by equation~(\ref{eq:SS_pF}) becomes:
\begin{equation}
	p(F)=p(F_n)=p(F_n|F_{n-1})p_0^{n-1}=p_0^{n-1}N_n/N
\end{equation}
where $N_n$ is the number of samples belonging to the last conditional level whose geocentric distance is lower than the Earth radius.

The main degrees of freedom of the method are the selected fixed conditional probability $p_0$, the number of samples per conditional level and the proposal auxiliary distribution for the MCMC phase, and they govern the accuracy and efficiency of the method \citep{Zuev2012}. We used for our analysis 1000 samples per conditional level and a value of conditional probability equal to 0.1. A normal distribution with spread equal to the original pdf was selected as proposal pdf for the MCMC algorithm.

A comparison between the SS technique and the ADP--IS method is shown in Table~\ref{tab:ADP-IS vs MC}. The required number of samples, the overall computational time, the estimated impact probability and related Poisson statistics uncertainty are shown. Results for the ADP--IS method are the ones referring to the $6\sigma$ case.

\begin{figure*}
	\includegraphics[width=0.7\textwidth]{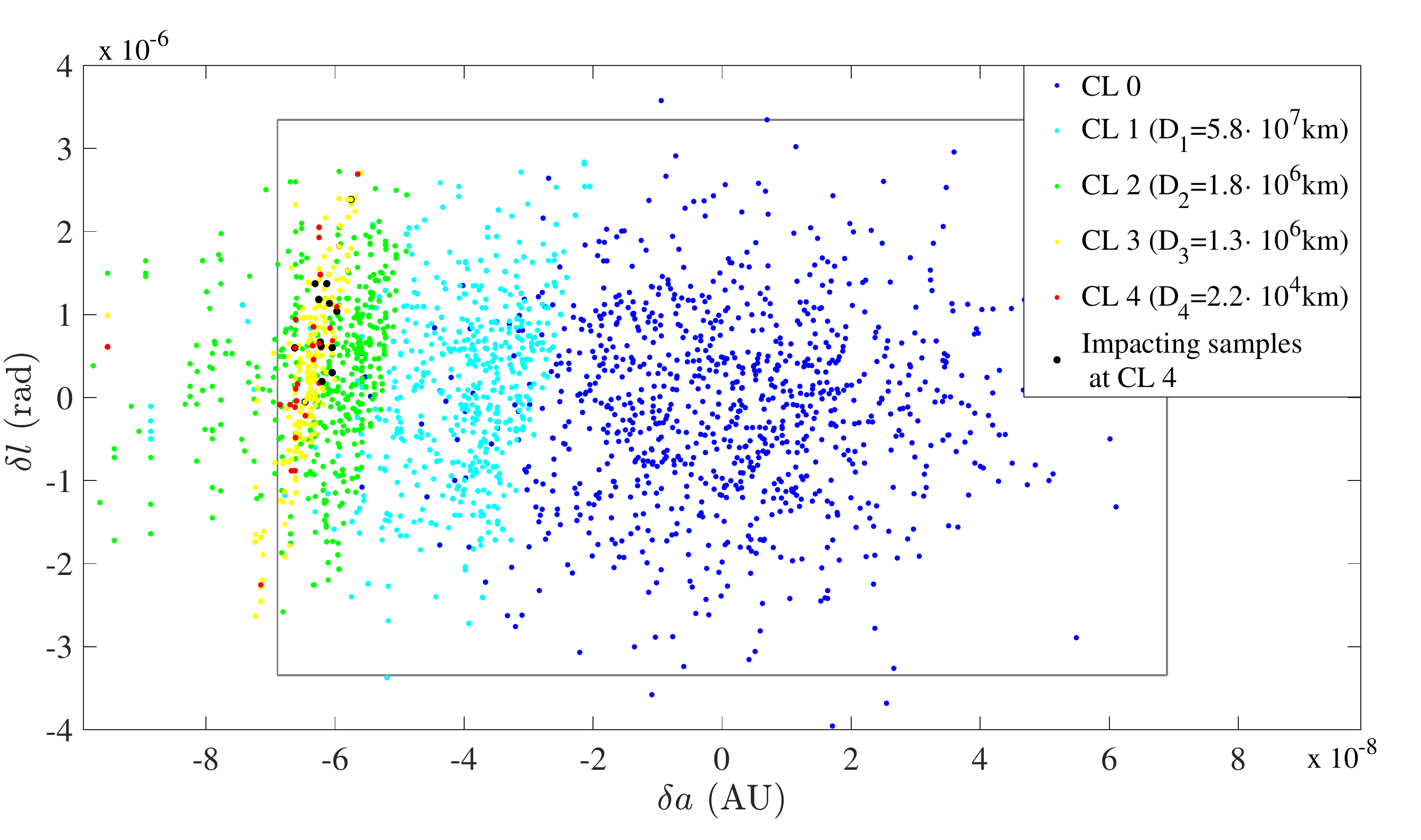}
    \caption{Subset Simulation conditional samples projected onto the $a-l$ plane ($N$ equal to 1000, $p_0$ equal to 0.1). In grey, boundaries of the $3\sigma$ domain.}
    \label{fig:Apophis_SS}
\end{figure*}

Subset Simulation and ADP--IS have a similar computational burden, though the required number of samples is very different. This result is expected, as the propagation windows for the two cases are different. Figure~\ref{fig:Apophis_SS} shows the distribution on the $a-l$ plane of the generated conditional samples obtained with SS, along with the thresholds per conditional level and related colors. Impacting samples at the last conditional level are represented in black. Conditional samples progressively move to the left, until impacting samples at conditional level 4 are identified. If compared to Figs.~\ref{fig:ADP_result_focus}-\ref{fig:ADP-IS_result}, this region is practically coincident with the PHS's identified during the ADP propagation. That is, SS and ADP--IS allow us to identify the same region in two completely independent ways. 

The advantage of the ADP--IS method is that, by identifying the PHS's, the propagation of the samples is drastically reduced in time, which yields a similar computational burden though the number of generated samples is significantly larger. Potentially, the ADP--IS method could take advantage of parallelization both during ADP propagation and the sampling phase, while the advantages for SS would be lower, as parallelization could be introduced only for specific phases of the algorithm. This approach would heighten the difference in efficiency between the two methods. However, the great savings granted by the SS could be included in the ADP--IS method during the sampling phase, by replacing the standard MC performed in the ISD with a SS limited to the unpruned subsets. This aspect may represent a future development of the method. Overall, the combination of ADP propagation and importance sampling allows us to achieve a computational burden that is competitive with both standard and advanced Monte Carlo techniques.

\begin{table*}
	\centering
	\caption{Comparison of the performance of ADP--IS method (order 5, tolerance $10^{-10}$, $N_{\text{max}}$ equal to 10, $6\sigma$ uncertainty set) and SS method ($N$ equal to 1000, $p_0$ equal to 0.1)}
	\label{tab:ADP-IS vs MC}
	\begin{tabular}{ccccc} 
		\hline
		 &  $n_{\text{samples}}$ & $t_{\text{all}}$ & $\hat{p}$ & $\sigma(\hat{p})$\\
		\hline
		ADP--IS & $353056$ & 1~h~28~min & $2.09\cdot 10^{-5}$ & $7.00\cdot 10^{-6}$\\
		
		SS & $4600$ & 1~h~30~min & $2.46\cdot 10^{-5}$ & $5.04\cdot 10^{-6}$\\
				
		\hline
	\end{tabular}
\end{table*}

Finally, it is worth comparing the performance of the presented approach with the reference technique in the field of impact probability computation, the LOV method. The LOV method takes advantage of the fact that the orbital uncertainty grows with time by stretching into a long slender ellipsoid in Cartesian space \citep{Farnocchia2015}. The tendency of uncertainty to stretch during propagation suggests the possibility of a one-dimensional parametrization of the uncertainty region, i.e. the sampling and the generation of the so-called Virtual Asteroids (VAs) is performed along the line of weakness of the orbit determination, and if all orbits are sufficiently close to the LOV, then significant savings in computational time with respect to a standard Monte Carlo approach are obtained without sacrificing reliability. 

The analysis presented in \citet{Milani2005} offers a first term of comparison: the generic completion level of $10^{-7}$ can be obtained with the propagation of only $\sim 10^{4}$ VAs. If compared to a standard MC approach, it would lead to compute times 3--4 orders of magnitude below those required for similar completeness with MC simulations \citep{Farnocchia2015}. The analysis presented in the previous section showed that the ADP--IS method grants a reduction in computation burden of around two orders of magnitude with respect to standard MC. Therefore, the LOV shows better performance than the ADS--IS method in the current implementation. 

Nevertheless, there are some cases in which the LOV method does not guarantee the same level of accuracy of a standard MC approach. A first case occurs when the observed arc of the investigated object is very short, i.e. 1 or 2 days \citep{Milani2005}. In this case, the confidence region is wide in two directions and the unidimensional sampling may not be suitable. What happens is that different LOVs, computed with different coordinates, provide independent sampling and may provide different results. That is, if some impacting samples lie well of the LOV and are separated from it by some strong nonlinearity, then the VAs selected along the LOV may fail to indicate some potential threatening encounters \citep{Milani2002}. In such cases, a standard MC approach would result more reliable, with unavoidable drawbacks in terms of computational time. As presented in this paper, the ADP--IS method, though maintaining a six-dimensional sampling, allows us to drastically reduce the computational effort by limiting the sampling to just specific regions. For these reasons, the method may be considered as a valuable trade-off between the efficiency of the LOV method and the reliability of standard MC in all those cases in which the former may result inaccurate. The possibility of improving the efficiency of the method by means of parallelization in both ADP propagation and sampling phases represents another step in this direction, as well as an optimised coding of the dynamics.

\section{Conclusions}

This paper introduced the combination of automatic domain pruning and importance sampling for uncertainty propagation and impact probability computation for Earth resonant returns of Near Earth Objects. The automatic domain pruning represents an evolution of the DA based automatic domain splitting technique, it allows us to estimate possible resonances after a planetary close encounter and limit the propagation of an uncertainty set to those subsets that may be involved in the resonant return of interest. During the propagation, the uncertainty domain is divided into subsets (Potentially Hazardous Subdomains) whose propagation stops just before the epoch of the resonant return. The identification of PHS's represents the starting point for the sampling phase. An importance sampling probability density function is defined over these subdomains and samples are drawn directly from this auxiliary pdf. We tested the ADP--IS method on the case of asteroid (99942) Apophis, providing an estimate for the impact probability in 2036. We carried out a sensitivity analysis on the main parameters of the method, providing general guidelines for their selection. The comparison with a standard Monte Carlo approach showed how the ADP--IS method can reduce the computation effort by more than two orders of magnitude, still granting the same accuracy level for the impact probability estimate. In addition, the current algorithm can be implemented to make use of parallelization techniques in both the ADP and the IS phase, thus significantly reduce the required computational time. All these considerations suggest that the method may be used as a valuable alternative to standard MC in all those cases in which the LOV method does not guarantee the required level of accuracy.
Future developments include a more rigorous formulation of the reference orbital period for subsets pruning allowing us to extend the pruning algorithm to the more critical case of intervening close encounters with other celestial bodies between the two encounters, and the testing to a wider set of cases.

\section*{Acknowledgements}
M. Losacco gratefully acknowledges professors E. Zio, N. Pedroni and F. Cadini from Politecnico di Milano for their introduction to the importance sampling and subset simulation techniques. In addition, the authors are grateful to the reviewer Davide Farnocchia for the constructive comments that significantly
improved the manuscript.
\newline
\newline
\textit{This is a pre-copyedited, author-produced PDF of an article accepted for publication in Monthly Notices of the Royal Astronomical Society following peer review. The version of record Matteo Losacco, Pierluigi Di Lizia, Roberto Armellin, Alexander Wittig; A differential algebra-based importance sampling method for impact probability computation on Earth resonant returns of near-Earth objects, Monthly Notices of the Royal Astronomical Society, Volume 479, Issue 4, 1 October 2018, Pages 5474-5490 is available online at: https://doi.org/10.1093/mnras/sty1832.}



\bibliographystyle{mnras}
\bibliography{Paper_ADP_IS_Losacco} 


\bsp	
\label{lastpage}
\end{document}